\documentclass[]{article}
\usepackage[margin=1.5in]{geometry}
\usepackage[english]{babel}
\usepackage[utf8x]{inputenc}
\usepackage{amsmath}
\usepackage{graphicx}
\usepackage{setspace}
\usepackage{lineno}
\usepackage{cite}

\title{Numerical Simulations of the Molecular Behavior and Entropy of Non-Ideal Argon}
\author{Matthew David Marko\\Marko Motors LLC\\mattdmarko@gmail.com}
\date{28 March 2026}

\begin{document}

\maketitle

\begin{abstract}
{\it 
A numerical model is built, simulating the principles of kinetic gas theory, to predict pressures of molecules in a spherical pressure vessel; the model tracks a single particle and multiplies the force on the spherical walls by a mole of molecules to predict the net pressure. An intermolecular attractive force is added for high-density simulations, to replicate a real fluid; the force is chosen to ensure the fluid matches the Peng-Robinson equation of state as it is compressed to a near supercritical density. The standard deviations of the molecule velocity with respect to temperature and density is studied to define the entropy. A parametric study of a Stirling cycle heat engine utilizing near-supercritical densities is modeled, to study how the temperature dependence of the attractive intermolecular Van der Waal forces can affect the net total entropy change to the surrounding environment. A practical, macroscopic-scale piston-cylinder engine was then built and demonstrated, utilizing a novel thermodynamic cycle that closely resembles the Carnot heat engine cycle, utilizing an arrangement of valves and pneumatic air to replicate the isothermal and isentropic compression and expansion of the working fluid. This heat engine cycle could be built without requiring advanced manufacturing, and utilized non-ideal carbon dioxide as the working fluid, to take advantage of the entropy effects of the Van der Waals forces demonstrated in the Argon simulations to boost the thermodynamic efficiency. This engine demonstrated this capability in a practical, macroscopic heat engine, and offers great opportunities to practical energy generation.
}
\end{abstract}

\section{Introduction}
\label{sec:Introduction}
In an earlier effort \cite{Marko_SciReports} from 2022, the author demonstrated experimental evidence that the intermolecular attractive Van der Waals forces can affect the overall generation of entropy in a heat engine cycle.  It was based on an analytical study \cite{MarkoAIP} from 2018, where a macroscopic Stirling cycle heat engine that utilized a non-ideal working fluid would have a net-negative entropy impact to the external universe.  This theory has been transferred to a working macroscopic heat engine system, that has been filed as a USPTO patent \cite{Marko_USPTO_HE_HP}, which is based on prior patents by the author \cite{Marko_USPTO_HE_simple,Marko_USPTO_HP}.  

The dimensionless thermodynamic efficiency $\eta$ of a heat engine is defined as the net work output $W_{OUT}$ (J) over the hot heat input $Q_H$ (J),
\begin{eqnarray}
\label{eq:eqEff_Thermo}
{\eta}&=&{\dfrac{W_{OUT}}{Q_H}}.
\end{eqnarray}
The maximum theoretical thermodynamic efficiency of an ideal-gas heat engine is limited by the Carnot efficiency ${\eta}_C$, 
\begin{eqnarray}
\label{eq:eqEffCarnot}
{\eta}_C&=&{1-{\frac{T_L}{T_H}}},
\end{eqnarray}
where $T_L$ (K) and $T_H$ (K) represent the low and high absolute temperatures.  In an ideal-gas heat engine cycle, the greater the temperature differential, the greater the maximum thermodynamic efficiency.  A full breakdown on the derivation of the Carnot efficiency ${\eta}_C$ is given in Appendix \ref{ap:appendix_Carnot}.  

It was demonstrated theoretically \cite{MarkoAIP} and experimentally \cite{Marko_SciReports} by the author that the change in internal energy of a real fluid undergoing isothermal expansion is simply the integration of the force necessary to overcome the intermolecular attractive Van der Waals forces \cite{ClausiusOrig,1,2,3,4,StatThermo,Keesom2, KeesomOrig, LondonDispOrig, IntermolCermanic, RS_DispersionTemp, TempDepPhysRevB, Entropy01, Entropy02, Entropy03, Entropy04, Entropy05, Entropy06, Carnot_PRL,Carnot_PRX,Carnot_NJOP,LJ_orig,sph,spam}.  This was realized by analyses of original measurements of the enthalpy of vaporization of water \cite{NISTsteamHvEqu, NISTsteamDataTable}; when calculating for the change in internal energy during vaporization, which is isothermal expansion of a real fluid, it was observed that the change in specific internal energy ${{\delta}u}_{{\delta}T=0}$ (J/kg) during isothermal compression and expansion followed a distinct empirical equation \cite{Marko_SciReports,MarkoAIP},
\begin{eqnarray}
\label{eq:eqDeltaU_isothermal_mytheory}
{{\delta}u}_{{\delta}T=0}&=&{{{a'}{\cdot}{{T}^{-0.25}}}{\cdot}{({\rho_1}-{\rho_2})}}, \\ 
{a'}&=&{\frac{0.21836}{9{\cdot}{({2^{\frac{1}{3}}-1})}}}{\cdot}{\frac{{R_G^2}{\cdot}{T_C^{2.5}}}{{P_C}}}.\nonumber 
\end{eqnarray}
where ${\rho}_1$ and ${\rho}_2$ (m$^3$/kg) represent the original and final density, \emph{T} (K) represents the absolute temperature, $R_G$ (J/kg${\cdot}$K) represents the gas constant, $T_C$ (K) represents the critical temperature, and $P_C$ (Pa) represents the critical pressure.  This equation has been found to match well for numerous fluids \cite{MarkoAIP,New_NIST_R134a,New_NIST_N2,New_NIST_H2O,New_NIST_CH4, New_NIST_C2H6, New_NIST_C3H8, New_NIST_C4H10n, New_NIST_C4H10iso, NISTargon1, NISTargon2, ArgonCV, ArgonHvThesis, ArgonCriticalProp, nistXn, Beattie_1951_Xn, HvNobelGases, nistN2, nistAmmonia, NISTsteamHvEqu, NISTwaterVolDat1, NISTwaterCritProp, NISTwaterVolDat2, NISTsteamDataTable, GoffGratch1946,nistCxHy,BWR1940}.  

To conform within Clausius' Theorem for the second law \cite{ClausiusOrig},
\begin{eqnarray}
\label{eq:eqClausius}
{{\oint}{\frac{{\delta}q}{T}}}&\leq&0,
\end{eqnarray}
the change in internal energy of a real fluid has previously been defined as, \cite{MarkoAIP,1,2,4}
\begin{eqnarray}
\label{eq:eqdU_ideal}
{\delta}{u}&=&{{C_V}{\cdot}{R_G}{\cdot}{{\delta}T}}+{\{{T{\cdot}{(\frac{{\partial}P}{{\partial}T}})}_V-{P}\}{\cdot}{{\delta}v}}, 
\end{eqnarray}
where $C_V$ is equal to the number of degrees of freedom of the molecule plus one half (ex. monatomic fluids $C_V=1.5$, diatomic fluids $C_V=2.5$, etc), ${\delta}T$ (K) represents the change in temperature, and ${\delta}v$ (m$^3$/kg) represents the change in specific volume.  Of note, the value of ${C_V}{\cdot}{R_G}$ (J/kg$\cdot$K) represents the specific heat capacity at a constant volume.  This derivation has led to negative internal energies in published thermodynamic tables, including the NIST Chemistry WebBook \cite{NIST_Webbook}, where the internal energy is arbitrarily set to zero at the triple-point.  A negative internal energy is obviously nonsensical.  

Another approach for determining non-isothermal changes in specific internal energy of a fluid ${{\delta}u}$ (J/kg) was demonstrated theoretically \cite{MarkoAIP} and experimentally \cite{Marko_SciReports} by the author.  First, one must include the specific intermolecular kinetic energy based off of kinetic theory $u_{KE}={C_V}{\cdot}{R_G}{\cdot}{T}$ (J/kg) \cite{KinTheoryBornGreen1946}; a detailed breakdown is given in Section \ref{sec:Kinetic_Theory}.  Next, by integrating the empirically derived equation for the change in internal potential energy (equation \ref{eq:eqDeltaU_isothermal_mytheory}) from a given density $\rho$ (kg/m$^3$) to infinitely low density (a true ideal gas) to find the intermolecular potential energy, and the temperature from absolute zero to the current temperature \emph{T}, one can calculate the total specific internal energy \emph{u} (J/kg) with equation \ref{eq:eqU_mytheory},
\begin{eqnarray}
\label{eq:eqU_mytheory}
{u}&=&{{C_V}{\cdot}{R_G}{\cdot}T}-{{{a'}{\cdot}{\rho}}{\cdot}{{T}^{-0.25}}},\\ \nonumber
{a'}&=&{{\frac{0.21836}{9{\cdot}{({2^{\frac{1}{3}}-1})}}}}{\cdot}{\frac{{R_G^2}{\cdot}{T_C^{2.5}}}{{P_C}}},
\end{eqnarray}
where \emph{T} (K) represents the absolute temperature, $R_G$ (J/kg${\cdot}$K) represents the gas constant, $\rho$ (kg/m$^3$) represents the density, $T_C$ (K) represents the critical temperature, $P_C$ (Pa) represents the critical pressure, and ${C_V}$ is equal to the number of degrees of freedom of the molecule plus one half (ex. monatomic fluids $C_V=1.5$, diatomic fluids $C_V=2.5$, etc).  This was realized experimentally \cite{Marko_SciReports} by rapidly expanding carbon dioxide \cite{CO2_NIST}, and taking measurements of the change in temperature as a result of the Joule-Thomson effect \cite{JouleThomson}; the measured internal energy of a real-fluid matched closer to equation \ref{eq:eqU_mytheory} than to the results in the NIST Chemistry WebBook \cite{NIST_Webbook} and equation \ref{eq:eqdU_ideal}.

In a theoretical Stirling or Carnot (Appendix \ref{ap:appendix_Carnot_HE_cycle}) heat engine thermodynamic cycle, the Van der Waals forces would both decrease the required work input during the cold isothermal compression, as well as reduce the work output recovered during the hot isothermal expansion.  These Van der Waals forces are observed to increase in strength with decreasing temperature; this is clearly described in existing empirical equations of state for real fluids, such as Redlich-Kwong \cite{RK1949} and Peng-Robinson \cite{PR1976,PitzerAcentric}.  Because of the temperature dependence of the Van der Waals forces, the reduction in cold work input is greater than the loss of the hot work output; therefore, the ideal efficiency of this macroscopic heat engine could in theory exceed the Carnot efficiency ${\eta}_C$ defined in equation \ref{eq:eqEffCarnot}.

An effort was made to numerically simulate a non-ideal working fluid in such a theoretical Stirling cycle heat engine.  A mole of argon was simulated, represented as an individual particle within a spherical container, traversing the container at many different angles until it impacted the container wall, causing pressure.  The pressure increase was not unlike the pressure expected of an ideal gas under Kinetic Theory (Section \ref{sec:Kinetic_Theory}), but the model was substantially modified to include an attractive intermolecular force; the details are described in depth in Section \ref{sec:Model_RF}.  A parametric study was then conducted at all the thermodynamic states of a non-ideal argon working fluid Stirling cycle heat engine, and the standard deviation of the molecule velocity was collected at each data point.  By looking at the data points, the simulation first demonstrated that the standard deviation of the molecular velocity has a clear ($>$0.99) correlation with the entropy of this fluid when the engine was effectively operating as an ideal gas.  Second, it is clear that the entropy (defined by this velocity-standard-deviation) is less for a non-ideal fluid subjected to intermolecular Van der Waals forces, and the increase in entropy of heating a real fluid can be less than the equivalent reduction in entropy from the ideal-gas ambient, leading to a net reduction in entropy to the ambient universe during this reversible cycle and a thermodynamic efficiency $\eta$ (equation \ref{eq:eqEff_Thermo}) that exceeds the Carnot efficiency $\eta_C$ (equation \ref{eq:eqEffCarnot}).  The full results of this numerical study is described in Section \ref{sec:Model_Results}.  The Fortran source code for this simulation is available in Appendix \ref{ap:appendix_fortran_code}.  

This effort seeks to both demonstrate this experimentally in a practical, macroscopic heat engine \cite{Marko_USPTO_HE_HP} that utilizes non-ideal carbon dioxide $CO_2$ \cite{CO2_NIST}.  This practical engine is actuated by a series of valves, connected to a pneumatic source, such that the carbon dioxide non-ideal working fluid approximates the Carnot cycle heat described in Appendix \ref{ap:appendix_Carnot_HE_cycle}, with a series of isothermal and isentropic compression and expansion.  The engine design is described in Section \ref{sec:Engine_Design}, and the thermodynamic analysis is described in Section \ref{sec:Engine_Analysis}.  

\section{Derivation of Kinetic Theory of an Ideal Gas}
\label{sec:Kinetic_Theory}
The kinetic model of an ideal gas \cite{2,3,KinTheoryBornGreen1946} is a well-established model to predict the kinetic energy of an ideal gas.  Internal energy, by definition, is the summation of the kinetic energy from all of the random molecular motion within a fluid, as well as any potential energy from intermolecular forces.  In the kinetic model, the gas is assumed to follow the ideal gas equation of state defined in equation \ref{eq:eqIdealGas}.  The ideal gas equation of state (equation \ref{eq:eqIdealGas}) can determine the pressure of an ideal gas, where the intermolecular Van der Waals forces are insignificant enough such that they can be disregarded,  
\begin{eqnarray}
\label{eq:eqIdealGas}
{P}&=&\dfrac{{R_G}{\cdot}{T}}{{v}},\\ \nonumber
{R_G}&=&{\dfrac{R_U}{MM}},\\ \nonumber
{R_U}&=&{{A_V}{\cdot}{k_B}},
\end{eqnarray}
where \emph{P} (Pa) represents the pressure, \emph{T} (K) represents the absolute temperature, \emph{v} (m$^3$/kg) represent the specific volume, $R_G$ (J/kg$\cdot$K) represents the gas constant, \emph{MM} (kg/mole) represents the molar mass, $R_U$ is the universal gas constant 8.314 (J/mole$\cdot$K), $A_V$ is Avogadro's Constant 6.022${\cdot}{10^{23}}$ mole$^{-1}$, and $k_B$ represents the Boltzmann Constant 1.380649 ${\cdot}{10^{-23}}$ J/K.  For the kinetic model to be applicable, the gas must be ideal, where all of the molecules are moving independent of each other, and there is no interaction between different gas molecules, either by collision or intermolecular forces \cite{3}.  

If a molecules is moving within the \emph{x} direction and hits the boundary of a container or pressure vessel, provided the gas is thermodynamically stable and there is no heat transfer, it will bounce off of the wall in the opposite direction.  The change in momentum for each molecular collision is therefore, 
\begin{eqnarray}
\label{eq:eqKT_01}
{\Delta}{p} = ({{m_m}{\cdot}{v_x}})-({-{m_m}{\cdot}{v_x}})={2{\cdot}{m_m}{\cdot}{v_x}},
\end{eqnarray}
where ${\Delta}{p}$ (kg$\cdot$m/s) is the change in momentum, ${m_m}$ (kg) is the mass of an individual molecule, and $v_x$ (m/s) is the velocity in the \emph{x}-direction.  The average time ${\Delta}{\tau}$ (s) for a molecule to cross the length \emph{L} (m) of the pressure vessel is,
\begin{eqnarray}
\label{eq:eqKT_02}
{\Delta}{\tau} &=& \frac{2{\cdot}L}{v_x}.  
\end{eqnarray}
The force applied to the walls of the pressure vessel with an individual molecular collision $F_m$ (Newtons) is the change in momentum per unit time, 
\begin{eqnarray}
\label{eq:eqKT_03}
F_m = \frac{{\Delta}{p}}{{\Delta}{\tau}}=\frac{{m_m}{\cdot}{v_x^2}}{L},
\end{eqnarray}
and the total force on the walls of the pressure vessel \emph{F} (Newtons) is thus, 
\begin{eqnarray}
\label{eq:eqKT_04}
F = {N}{\cdot}{F_m}=\frac{{N}{\cdot}{m_m}{\cdot}{v_x^2}}{L},
\end{eqnarray}
where \emph{N} is the total count of the molecules.  

So far this analysis has only been in the \emph{x}-direction, when in reality the molecules are bouncing in three dimensions.  Assuming the average speed in all three directions are identical, as is the case in a stable fluid, according to Pythagorean theorem the average Root Mean Square (RMS) total velocity ${v_{RMS}}$ (m/s) is thus, 
\begin{eqnarray}
\label{eq:eqKT_05}
{v_{RMS}^2} = {v_x^2}+{v_y^2}+{v_z^2} = {3}{\cdot}{v_x^2},
\end{eqnarray}
and thus equation \ref{eq:eqKT_04} can be rewritten as, 
\begin{eqnarray}
\label{eq:eqKT_06}
F &=& \frac{{N}{\cdot}{m_m}{\cdot}{v_{RMS}^2}}{3{\cdot}L}. 
\end{eqnarray}
In the kinetic theory, equation \ref{eq:eqKT_06} would only apply to molecules that have no rotational or vibrational energies, specifically monatomic molecules such as helium, neon, argon, xenon, krypton, or radon gas \cite{3}.  

The pressure, by definition, is merely the ratio of the total force over the area of the container, and therefore assuming the container is cubic in shape, the pressure \emph{P} (Pa) is, 
\begin{eqnarray}
\label{eq:eqKT_07}
P = \frac{F}{L^2}= \frac{{N}{\cdot}{m_m}{\cdot}{v_{RMS}^2}}{3{\cdot}L^3}=\frac{{N}{\cdot}{m_m}{\cdot}{v_{RMS}^2}}{3{\cdot}V}, 
\end{eqnarray}
where \emph{V} ($m^3$) is the volume of the container.  

The total kinetic energy of the gas \emph{KE} (J) is defined as the sum of the kinetic energies of the gas molecules, 
\begin{eqnarray}
\label{eq:eqKT_08}
KE&=&{\frac{1}{2}}{\cdot}N{\cdot}{m_m}{\cdot}{v_{RMS}^2},
\end{eqnarray}
and therefore plugging equation \ref{eq:eqKT_08} into equation \ref{eq:eqKT_07},
\begin{eqnarray}
P &=& \frac{{2}{\cdot}{KE}}{3{\cdot}V},\nonumber
\end{eqnarray}
and therefore the kinetic energy of a monatomic ideal gas can be defined as, 
\begin{eqnarray}
\label{eq:eqKT_09}
KE&=&{\frac{3}{2}}{\cdot}{P}{\cdot}{V}.  
\end{eqnarray}
As the kinetic model is dealing with an ideal gas, equation \ref{eq:eqIdealGas} is applicable, and thus \cite{3}, 
\begin{eqnarray}
\label{eq:eqKT_10}
KE={\frac{3}{2}}{\cdot}{P}{\cdot}{V}={\frac{3}{2}}{\cdot}{m_T}{\cdot}{R_G}{\cdot}{T}={\frac{3}{2}}{\cdot}{N}{\cdot}{{k_B}}{\cdot}{T}, 
\end{eqnarray}
where $m_T$ (kg) is the total mass of the gas, 
\begin{eqnarray}
{m_T}&=&N{\cdot}{m_m}.\nonumber
\end{eqnarray}
The relationship between temperature and kinetic energy is thus defined with equations \ref{eq:eqKT_08} and \ref{eq:eqKT_10}.  This can be rewritten as, 
\begin{eqnarray}
KE={\frac{1}{2}}{\cdot}N{\cdot}{m_m}{\cdot}{v_{RMS}^2}={\frac{3}{2}}{\cdot}{N}{\cdot}{k_B}{\cdot}{T},\nonumber
\end{eqnarray}
and thus the average total velocity $v_{RMS}$ (m/s) of a particle of an ideal gas is proportional to the square root of the absolute temperature \emph{T} (K) \cite{3}, 
\begin{eqnarray}
\label{eq:eqKT_11}
v_{RMS}=\sqrt{\frac{{3}{\cdot}{k_B}{\cdot}{T}}{m_m}}.
\end{eqnarray}

\section{Modeling and Simulation of a non-ideal fluid}
\label{sec:Model_RF}
\subsection{Kinetic Simulation of ideal-gas molecules}

A model was build in the Fortran programming language, to simulate one mole (6.02214086${\cdot}{10^{23}}$) of argon molecules traveling in a spherical volume.  Argon was chosen because it is a simple monatomic molecule, commonly used in industry, and its critical properties are not at excessively low temperatures (ex. Helium).  Argon has a molar mass \emph{MM} of 39.9 g/mole, a critical pressure $P_C$ of 4.863 MPa, a critical temperature ${T_C}$ of 150.687 K, a critical density $\rho_C$ of 535 kg/m$^3$, and a critical specific volume $V_C$ of 1.8692 cm$^3$/g \cite{ArgonCriticalProp}.  

The model will take the dimensionless reduced temperature $T_R$ and reduced specific volume $V_R$ as inputs, 
\begin{eqnarray}
\label{eq:eqReduced_Tr_Vr}
T_R&=&\dfrac{T}{T_C},\\ \nonumber
v_R&=&\dfrac{v}{v_C},
\end{eqnarray}
where \emph{T} (K) is the absolute temperature, and \emph{v} (m$^3$/kg) is the specific volume.  The absolute temperature \emph{T} (K) is easily calculated as $T={T_R}{\cdot}{T_C}$, and the volume (for one mole) is calculated as ${V_{sphere}}={V_R}{\cdot}{V_C}{\cdot}{MM}$.  From the known volume of the sphere, the radius and surface area are easily calculated as, 
\begin{eqnarray}
\label{eq:eqRandAsphere}
{R_{sphere}}&=&{({{\frac{3}{{\pi}{\cdot}4}}{\cdot}{V_{sphere}}})}^{\frac{1}{3}},  \\ \nonumber 
{A_{sphere}}&=&{4}{\cdot}{\pi}{\cdot}{R_{sphere}^2},
\end{eqnarray}

The model has the option of simulating the particle at a constant speed for a given temperature, if so the speed is constantly the $v_{RMS}$ speed for the given temperature defined in equation \ref{eq:eqKT_11}.  The model also gives the option of simulating a profile of faster and slower speeds; the speed profile will maintain the same \emph{RMS} average speed defined in equation \ref{eq:eqKT_11}, and the average speed ${v_{avg}}$ (m/s) will be determined as, 
\begin{eqnarray}
\label{eq:eqAvgRMSrat}
{v_{avg}}&=&{v_{RMS}}{\cdot}{\sqrt{\frac{8}{3{\cdot}{\pi}}}}.
\end{eqnarray}
If the model calls for $N_Y$ velocity increments to be simulated, a subroutine in the Fortran code will generate a ${N_Y}{\cdot}{1}$ vector-array, ranging from 0.2 to 1.8, averaging 1.0, with a standard deviation of 0.71.  This vector-array will be multiplied by the average molecule velocity at the boundary ${v_{avg}}$; the \emph{RMS} of the velocity vector-array will be equal to ${v_{RMS}}$ determined with \ref{eq:eqKT_11}.  Within this simulation, a value of $N_Y$ of 100 is used.  

The time-step ${\delta}{\tau}$ (s) is determined by the estimated time for an argon molecule traveling at the average speed $v_{avg}$ (m/s) across the diameter of the sphere $2{\cdot}{R_{sphere}}$ (m).  This time is divided by the integer value $N_{{\delta}{\tau}}$ that is specified by the model, to give a time-step,  
\begin{eqnarray}
{{\delta}{\tau}}={({\frac{2{\cdot}{R_{sphere}}}{v_{avg}}})}{\cdot}{\frac{1}{N_{{\delta}{\tau}}}}.
\end{eqnarray}
It is necessary to record the molecule's position and velocity with each increment, but with different angles and speeds, it is impossible to know exactly how many time steps will be needed for each test parameter.  In this Fortran code, an array length of ${10}{\cdot}{N_{{\delta}{\tau}}}$ was found to be more than enough to avoid any risk of running out of array space.  In this study, a resolution of ${N_{{\delta}{\tau}}}$=300 was used; increasing the resolution beyond this number was not observed to have any significant impact on the results.  

At each velocity increment, the model simulates a molecule leaving the surface of the sphere at different angles.  As a sphere is effectively identical at all surface locations, the point of initial contact will be defined as \emph{(-R,0,0)}.  The initial velocity will be defined in three dimensions as, 
\begin{eqnarray}
\label{eq:eqVxyzFct}
{V_x}&=&{V_i}{\cdot}{sin(\theta)}{\cdot}{cos(\phi)} \\
{V_y}&=&{V_i}{\cdot}{sin(\theta)}{\cdot}{sin(\phi)} \nonumber \\
{V_z}&=&{V_i}{\cdot}{cos(\theta)}, \nonumber 
\end{eqnarray}
where $\phi$ ranges from 0 to ${\pi}/2$, and $\theta$ ranges from 0 to $\pi$, both in 91 increments, resulting in 91$^2$=8,281 different directions.  The velocity magnitude $V_i$ (m/s) for the individual increment is determined from the temperature (equation \ref{eq:eqKT_11}), and $N_Y$=100; there is thus a total of 828,100 simulations for each temperature and volume increment.  

The kinetic gas theory assumes the molecule travels across the long length of the volume and directly impacts the wall; in reality molecules will travel at all possible angles.  If a molecule were to travel directly through the center of the sphere, the time ${\tau}$ (s) to travel will simply be ${\tau}={2{\cdot}{R_{sphere}/{v_{avg}}}}$, and the force due to the change in momentum for a single molecule will be derived from equation \ref{eq:eqKT_06}, where $F={{m_m}{\cdot}{v_{RMS}^2}}/({3{\cdot}2{\cdot}{R_{sphere}}})$.  Assuming the spherical volume, if a molecule were to travel at an angle from the center of the sphere, the travel time $\tau$ (s) will be reduced, but the force will also be reduced as the molecule is hitting the surface at an angle, and will only transmit part of its energy to changing momentum and direction.  

The simulation starts off with a molecule at position (-R,0,0).  With each time-step, it increments the three dimensions based on the 3-dimensional velocity described in equation \ref{eq:eqVxyzFct}.  The model uses a \emph{while} loop until the radius ${r_{ii}}$ (m) of the position, 
\begin{eqnarray}
\label{eq:eqRadiusXYZ}
{r_{ii}}&=&\sqrt{{x^2}+{y^2}+{z^2}},
\end{eqnarray}
exceeds the radius of the sphere, ${r_{ii}}>{R_{sphere}}$.  At this point, the molecule has impacted the cylinder wall.  If the molecule travels right through the center and impacts the other end at position (R,0,0), then the velocity will be ${V_i}{\cdot}(1,0,0)$, and the force impacted will be at a maximum; the travel time $\tau$ (s) will also be the maximum ${\tau}={2{\cdot}{R_{sphere}/{V_{i}}}}$.  If the molecule were to travel at a 90$^{\circ}$ perpendicular direction, where the velocity were ${V_i}{\cdot}(0,1,0)$ or ${V_i}{\cdot}(0,0,1)$, the position will remain at (-R,0,0) and the travel time $\tau$ will effectively be 0.  For all the molecules traveling at angles in between the two extremes, the force applied is simply the dot product of the velocity with the position of the impact, 
\begin{eqnarray}
\label{eq:eqDotProdRat}
VX_{rat}&=&{{X_x}{\cdot}{V_x}}+{{X_y}{\cdot}{V_y}}+{{X_z}{\cdot}{V_z}},
\end{eqnarray}
and the dimensionless $VX_{rat}$ is applied to the equation for the force applied by a single molecule ${F_m}$ (N) defined in equation \ref{eq:eqF_diffAng}, 
\begin{eqnarray}
\label{eq:eqF_diffAng}
{F_m}&=&\frac{{VX_{rat}}{\cdot}{m_m}{\cdot}{\sqrt{{V_x^2}+{V_y^2}+{V_z^2}}}}{{3{\cdot}2{\cdot}{R_{sphere}}}}.  
\end{eqnarray}

Throughout this simulation, for all initial angles and velocities, the position and velocity in three dimensions is tabulated and recorded.  Each position and velocity is stored in a large data file, and at the conclusion of the simulation, the average, RMS, and standard deviation of both the positions and the velocities are determine.  The purpose of determining the standard deviation is to find the relationship between the standard deviation of the velocity as it relates to entropy (equation \ref{eq:eqSideal}), where the change in specific entropy ${\delta}s$ (J/kg$\cdot$K) is defined as \cite{1,2,3,4,StatThermo}, 
\begin{eqnarray}
\label{eq:eqSideal}
{\delta}{s}&=&\frac{q}{T},
\end{eqnarray}
where \emph{T} (K) is the absolute temperature, and \emph{q} (J/kg) represent the heat transferred per unit mass.

\subsection{Kinetic-Potential Simulation}

As the density of a fluid increases to the point of being a saturated liquid, saturated gas, or supercritical fluid, intermolecular attractive (and repulsive) forces \cite{KeesomOrig, Keesom2, LondonDispOrig, IntermolCermanic, RS_DispersionTemp, TempDepPhysRevB} can impact the pressure and temperature of the fluid.  For a real fluid, the ideal gas equation of state (equation \ref{eq:eqIdealGas}) no longer is applicable, and one needs to use an empirically-derived equation of state that takes into account the intermolecular Van der Waals forces, such as the Peng-Robinson equation of state defined in equation \ref{eq:eqPengRobinson}.  As the molecules get closer together in the presence of attractive intermolecular forces, the internal potential energy will decrease.  

The ideal gas equation of state (equation \ref{eq:eqIdealGas}) breaks down in the presence of intermolecular attractive and repulsive Van der Waals forces, and therefore empirical equations of states are used, such as the Redlich-Kwong \cite{RK1949} and the Peng-Robinson equation of state \cite{PR1976,PitzerAcentric} defined in equation \ref{eq:eqPengRobinson},
\begin{eqnarray}
\label{eq:eqPengRobinson}
{P}&=&{\frac{{R_G}{\cdot}T}{v-B}}-{\frac{A{\cdot}{\alpha}}{{v^2}+{2{\cdot}B{\cdot}v}-{B^2}}},\\ \nonumber
A&=&{{\Omega}_A}{\cdot}{\frac{{R^2}{\cdot}{T_C^2}}{P_C}}, \\ \nonumber
B&=&{{\Omega}_B}{\cdot}{\frac{R{\cdot}{T_C}}{P_C}}, \\ \nonumber
{\Omega}_A&=&{\frac{8+{40{\cdot}{\Omega_C}}}{49-{37{\cdot}{\Omega_C}}}}=0.45724 \\ \nonumber
{\Omega}_B&=&\frac{{\Omega_C}}{{\Omega_C}+3}=0.07780\\ \nonumber
{\Omega}_C&=&\frac{1}{1+{{(4-\sqrt{8}})^{1/3}}+{{(4+\sqrt{8}})^{1/3}}}=0.25308;\\ \nonumber
{\alpha}&=&{({1+{{\kappa}{\cdot}{({1-{\sqrt{T_R}}})}}})^2}, \\ \nonumber
{\kappa}&=&{0.37464}+{1.54226{\cdot}{\omega}}-{0.26992{\cdot}{\omega^2}},\nonumber
\end{eqnarray}
where $\omega$ is Pitzer's acentric factor \cite{PitzerAcentric}, defined as
\begin{eqnarray}
\label{eq:eqPitzerAcentricFactor}
{\omega}&=&log_{10}(\frac{P_C}{P_{S}'})-1,
\end{eqnarray}
where $P_{S}'$ (Pa) is the saturated pressure at a reduced temperature of $T_R=T/{T_C}=0.7$, and $P_C$ (Pa) is the critical pressure.  For all of the monatomic fluids including argon, ${\omega}=0$.  The coefficient \emph{A} represents the intermolecular attractive force, and the coefficient \emph{B} represents the actual volume of the molecules at absolute zero.  As the specific volume \emph{v} (m$^3$/kg) increases (and the density decreases), equation \ref{eq:eqPengRobinson} matches the ideal gas law defined in equation \ref{eq:eqIdealGas}.  

If dealing with a purely ideal gas, molecules have no interaction with each other, and the pressure and velocities can be solved with the purely analytical approach of the kinetic gas theory.  To model real fluids, with intermolecular Van der Waals fluids, assumptions for the intermolecular forces are necessary.  In Lennard Jones' equation, the attractive Van der Waals force ${F_{VDW}}$ (N) for two molecules is proportional to the distance between particles to the sixth exponent \cite{4,LJ_orig}, 
\begin{eqnarray}
\label{eq:eqVDWforcesGeneral}
{F_{VDW}}=\frac{a'}{r^6},
\end{eqnarray}
where \emph{a'} is a constant and \emph{r} (m) is the distance between two molecules.  While the Lennard Jones potential equation \cite{LJ_orig} also includes a twelfth power for the repulsive forces, these are not based in reality, and the repulsive forces due to the Pauli Exclusion Principle are considered by subtracting the minimum possible volume \emph{B} (m$^3$/kg) in the equation of state such as in Peng-Robinson equation \ref{eq:eqPengRobinson}.  

For the sake of simplicity, assume that the volume is a perfect sphere of a real, monatomic fluid molecules following the Peng-Robinson equation of state.  The surface area ${A_{sphere}}$ (m$^2$) and volume of this sphere ${V_{sphere}}$ (m$^3$) is simply, 
\begin{eqnarray}
\label{eq:eqVolSphere}
{V_{sphere}}={\frac{4}{3}}{\cdot}{\pi}{\cdot}{R_{sphere}^3},\\ \nonumber
{A_{sphere}}={4}{\cdot}{\pi}{\cdot}{R_{sphere}^2},
\end{eqnarray}
where ${R_{sphere}}$ (m) represents the sphere radius.  Next, assume a molecule is on the far edge of this sphere; to determine the net attractive forces one must determine the summation of the average distances of the other molecules within the volume,  
\begin{eqnarray}
\label{eq:eqProbArea01}
{\hat{P}(x)}=\frac{A(x)}{A_{avg}}.
\end{eqnarray}
The cross-section area of the sphere at a given \emph{X}-axis point \emph{A(x)} can be found from the radius of the cross section, 
\begin{eqnarray}
\label{eq:eqAxfct}
{A(x)}&=&{\pi}{\cdot}{R_{sphere}^2}{\cdot}{cos^2({sin^{-1}(\frac{x}{R_{sphere}})})},
\end{eqnarray}
while the average cross section area is simply the total volume of the sphere over the diameter of the sphere, 
\begin{eqnarray}
\label{eq:eqAavg}
{A_{avg}}&=&\frac{{\frac{4}{3}}{\cdot}{\pi}{\cdot}{R_{sphere}^3}}{2{\cdot}{R_{sphere}}},\\ \nonumber
&=&{\frac{2}{3}}{\cdot}{\pi}{\cdot}{{R_{sphere}}^2},
\end{eqnarray}
and now the probability ${\hat{P}(x)}$ can be found by plugging the results of equation \ref{eq:eqAxfct} and \ref{eq:eqAavg} into equation \ref{eq:eqProbArea01}, 
\begin{eqnarray}
\label{eq:eqProbAreaFinal}
{\hat{P}(x)}={\frac{3}{2}}{\cdot}{cos^2({sin^{-1}(\frac{x}{R_{sphere}})})}. 
\end{eqnarray}

The next step is to integrate across the diameter of the sphere along the \emph{X}-axis in order to find the overall average distance to the sixth power ${\bar{{\delta}{x}^6}}$ (m), 
\begin{eqnarray}
{\bar{{\delta}{x}^6}}&=&{{\int}_{-R}^R}{(R-x)^6}{\cdot}{\hat{P}(x)}dx,\\
&=&{{\int}_{-R_{sphere}}^{R_{sphere}}}{({R_{sphere}}-x)^6}{\cdot}{{\frac{3}{2}}{\cdot}{cos^2({sin^{-1}(\frac{x}{R_{sphere}})})}}dx,\nonumber \\ \nonumber
&=&{\frac{16}{3}}{\cdot}{{R_{sphere}}^6}. 
\end{eqnarray}
It is desired not just for the average distance to a particle at the edge of the sphere, but all throughout the radius.  A particle moving on the \emph{X}-axis will experience attraction from particles both in front of and behind it, and therefore the proper average ${\bar{{\delta}{x}^6}}$, for the purpose of determining net total attraction towards the center of the sphere, 
\begin{eqnarray}
\label{eq:eqDX6full}
{\bar{{\delta}{x}^6}}(r)&=&{{{\int}_{-{R_{sphere}}}^r}{(r-x)^6}{\cdot}{{\frac{3}{2}}{\cdot}{cos^2({sin^{-1}(\frac{x}{R_{sphere}})})}}dx}-{\ldots}\\ \nonumber
&&{\ldots}{{{\int}_{r}^{R_{sphere}}}{({R_{sphere}}-x)^6}{\cdot}{{\frac{3}{2}}{\cdot}{cos^2({sin^{-1}(\frac{x}{R_{sphere}})})}}dx},
\end{eqnarray}
which can be simplified by the approximate equation \ref{eq:eqDX6simple}, 
\begin{eqnarray}
\label{eq:eqDX6simple}
{\bar{{\delta}{x}^6}}(r)&{\approx}&{\frac{16}{3}}{\cdot}{r^3},
\end{eqnarray}
where \emph{r} (m) represents the radial position on the \emph{X}-axis, where $0<r<R_{sphere}$.  The correlation coefficient between the two equations \ref{eq:eqDX6full} and \ref{eq:eqDX6simple}, where \emph{$\delta$r=0.001}, is R = 0.99936; the results are tabulated in Table \ref{tb:tbX6}. 

\begin{table}[h]
\begin{center}
\begin{tabular}{ | c || c | c | }
  \hline
r & ${\bar{{\delta}{x}^6}}(r)$ & ${\bar{{\delta}{x}^6}}(r)$ \\
 & eq. \ref{eq:eqDX6full} & eq. \ref{eq:eqDX6simple} \\
  \hline
0.05005 & 0.033158 & 0.00066867\\  
0.1001 & 0.06865 & 0.0053494\\  
0.15015 & 0.10887 & 0.018054\\  
0.2002 & 0.15633 & 0.042795\\  
0.25025 & 0.21373 & 0.083584\\  
0.3003 & 0.28399 & 0.14443\\  
0.35035 & 0.37035 & 0.22935\\  
0.4004 & 0.4764 & 0.34236\\  
0.45045 & 0.60615 & 0.48746\\  
0.4995 & 0.76063 & 0.66467\\  
0.54955 & 0.95109 & 0.88515\\  
0.5996 & 1.1803 & 1.1497\\  
0.64965 & 1.4545 & 1.4623\\  
0.6997 & 1.7809 & 1.827\\  
0.74975 & 2.1671 & 2.2477\\  
0.7998 & 2.6218 & 2.7286\\  
0.84985 & 3.1547 & 3.2736\\  
0.8999 & 3.7763 & 3.8867\\  
0.94995 & 4.4982 & 4.5719\\  
1 & 5.3333 & 5.3333\\  
  \hline
\end{tabular}
\caption{Comparison of ${\bar{{\delta}{x}^6}}(r)$ functions between equation \ref{eq:eqDX6full} and \ref{eq:eqDX6simple}.  The correlation coefficient between the two equations ($\delta$r=0.001) is R = 0.99936. }
\label{tb:tbX6}
\end{center}
\end{table}

While typical conservative forces such as gravity, electrostatic forces, and Van der Waals attractive forces increase as the distance between two attractive objects decreases, it is clear from equation \ref{eq:eqDX6full} that the forces will decrease when a given molecule moves closer to the center of the volume, proportional to the radial position cubed.  This makes physical sense, as near the center of the sphere, the attractive forces of neighboring molecules on one side of the molecule counteract the attractive forces from the other side.  

When modeling the effects of intermolecular attractive forces, it is not enough to simply take the pressure reduced from the intermolecular attractive Van der Waals force, multiply it by the spherical surface area, divide it by the number of molecules, and reduce it by the relative radius cubed.  The reason for this is that the overall change in pressure of the real fluid includes the pressure reduced from the attractive force, as well as the change in time for the molecule to travel across the spherical volume.  An increasing force will inherently accelerate the molecule towards the center, and decelerate it towards the other side, reducing the travel time, and thus increasing the pressure.  It is necessary to select a force that balances these two impacts on the final pressure, in order to achieve the correct pressure for the equation of state.  

A parametric study of the supercritical argon molecules propagating in the sphere was conducted to determine the exact function for the intermolecular force on each molecule.  The maximum such a force will be is that which will cause the drop in pressure observed in most empirical equations of states, such as the Peng-Robinson defined in equation  \ref{eq:eqPengRobinson}, 
\begin{eqnarray}
\label{eq:eq_dP_PengRobinson}
{\delta}{P}&=&-{\frac{A{\cdot}{\alpha}}{{v^2}+{2{\cdot}B{\cdot}v}-{B^2}}}.
\end{eqnarray}
The derivative of the change in internal energy defined in equation \ref{eq:eq_dP_PengRobinson} (empirically) gives a very close approximation for ${\delta}{P}$, 
\begin{eqnarray}
{\delta}{P}&\approx&-{\frac{{R^2}{\cdot}{T_C^{2.5}}}{9{\cdot}{({2^{\frac{1}{3}}-1})}{\cdot}{P_C}}}{\cdot}{{\frac{1}{\sqrt{T}}}{\cdot}{{\frac{1}{v^2}}}}. \nonumber 
\end{eqnarray}
The force needs to be some ratio of this, as increasing the force will increase the average speed of a molecule (for a given $v_{RMS}$ (m/s) at the surfaces), reducing the time in between impacted the sphere's surface, and increasing the pressure.  

A parametric study was performed to find the exact ratio of this pressure, and a function for the force (N) on a given molecule, accelerating it as it travels towards the center and decelerating it as it travels back towards the surface, was determined in order that the molecule satisfy the Peng-Robinson equation of state defined in equation \ref{eq:eqPengRobinson}.  The Van der Waals attractive force $F_{VDW}$ (N) is thus, 
\begin{eqnarray}
\label{eq:eqFvdw}
F_{VWD}&=&{{\chi}}{\cdot}{{\frac{{R^2}{\cdot}{T_C^{2.5}}}{9{\cdot}{({2^{\frac{1}{3}}-1})}{\cdot}{P_C}}}{\cdot}{{\frac{1}{\sqrt{T}}}{\cdot}{{\frac{1}{v^2}}}}}{\cdot}{\frac{A_{sphere}}{N}},
\end{eqnarray}
where \emph{N} is the number of molecules in the sphere (one mole for this simulation), and $\chi$ is a dimensionless coefficient, 
\begin{eqnarray}
\label{eq:eqFvdw_Coeff}
{\chi}=&2.3246-{\frac{0.8441}{\sqrt(V_R)}}-{{0.8670}{\cdot}{T_R}},& {T_R}{\leq}1, \\ \nonumber
=&2.3246-{\frac{0.8441}{\sqrt(V_R)}}-{{0.8670}{\cdot}{\sqrt{T_R}}},& {T_R}{\geq}{1}, 
\end{eqnarray}
determined from the parametric study.  This force $F_{VDW}$ (N) is the force at the surface of the sphere towards the center; this force decreases in each of the three dimensions as the molecule gets closer to the center, 
\begin{eqnarray}
F_x&=&-{F_{VDW}}{\cdot}{(\frac{x}{R})^3},\nonumber \\
F_y&=&-{F_{VDW}}{\cdot}{(\frac{y}{R})^3},\nonumber \\
F_z&=&-{F_{VDW}}{\cdot}{(\frac{z}{R})^3}.\nonumber 
\end{eqnarray}

\section{Modeling of the Impacts of the intermolecular attractive Van der Waals forces on entropy}
\label{sec:Model_Results}
\subsection{Defining entropy as molecular velocity standard deviation}

A parametric simulation of this model was conducted, with twenty temperature parameters and ten volume parameters, all with one mole of argon held in a sphere, for a total of 200 independent simulations of 828,100 molecular simulations.  The reduced temperature and reduced volume are defined as follows, 
\begin{eqnarray}
T_R&=&exp((ii-1){\cdot}0.1),\nonumber \\
V_R&=&exp((jj-1){\cdot}0.25),\nonumber 
\end{eqnarray}
where \emph{ii} ranges from 1 to 20, and where \emph{jj} ranges from 1 to 10.  The actual absolute temperature \emph{T} (K) of the argon is simply $T={T_R}{\cdot}{T_C}$, where the critical temperature ${T_C}$ of argon is 150.687 K \cite{ArgonCriticalProp}.  The actual volume of the sphere $V_{sphere}$ (m$^3$) for one mole of argon was determined from the reduced volume $V_R$, the specific density $\rho_C$=535 kg/m$^3$, and the molar mass \emph{MM}=39.9 g/mole simply by, 
\begin{eqnarray}
\label{eq:eqVol_VrVcMM}
V_{sphere}&=&\frac{{V_R}{\cdot}{MM}}{\rho_C},
\end{eqnarray}
and thus the radius $R_{sphere}$ (m) and surface area $A_{sphere}$ (m$^2$) of the sphere can be determined from $V_{sphere}$ (m$^3$) with equation \ref{eq:eqRandAsphere}.  In the parametric simulation, the numerical prediction for the pressure matched the Peng-Robinson $P_{PR}$ (equation \ref{eq:eqPengRobinson}) pressures within less than 5\% error, and the correlation \emph{R} between the simulated results and Peng-Robinson equation is 0.990.  

After the numerical simulations were completed, the standard deviation of the velocity ${\sigma}_{V}$ (m/s) was made dimensionless by dividing it by the critical temperature squared, 
\begin{eqnarray}
\label{eq:eqSTD_V_norm}
{\Bar{\sigma}_{V}}&=&\frac{{{\sigma}_{V}}}{T_C^2}.  
\end{eqnarray}
All of the data points that constituted an ideal gas were segregated; an ideal gas was one that the pressure with the Peng-Robinson equation of state (equation \ref{eq:eqPengRobinson}) matched the ideal gas pressure equation of state (equation \ref{eq:eqIdealGas}) within 5\%.  Using regression analysis, an equation for the normalized velocity standard deviation ${\Bar{\sigma}_{V}}$,
\begin{eqnarray}
\label{eq:eqSTD_V_norm_regression_IG}
{\Bar{\sigma}_{V,IG}}&=&{\{0.6118+{0.9336{\cdot}{log(T_R)}}+{0.0471{\cdot}{log(V_R)}}\}}^2. 
\end{eqnarray}
The correlation between the regression analysis equation \ref{eq:eqSTD_V_norm_regression_IG} and the data was 0.9960, the mean average error was 1.46\%, and the median error was 1.34\%.  

Next, all of the data points of a real fluid were segregated; a real fluid was one that the pressure with the Peng-Robinson equation of state (equation \ref{eq:eqPengRobinson}) deviated from the ideal gas pressure equation of state (equation \ref{eq:eqIdealGas}) by more than 10\%.  The difference between the simulated normalized velocity standard deviation ${\Bar{\sigma}_{V}}$ and the ideal-gas normalized velocity standard deviation defined in equation \ref{eq:eqSTD_V_norm_regression_IG} can be realized, 
\begin{eqnarray}
\label{eq:eqSTD_V_norm_regression_error_RF}
{\sqrt{\Bar{\sigma}_{V,IG}}}-{\sqrt{\Bar{\sigma}_{V}}}&=&{\{0.6299+{0.3085{\cdot}{T_R}}+{8.9709{\cdot}{10^{-3}}{\cdot}{V_R}}\}}. 
\end{eqnarray}
It can be clearly observed in equation \ref{eq:eqSTD_V_norm_regression_error_RF} that the numerical simulations demonstrate that there is a reduction in the standard deviation of the velocity ${\Bar{\sigma}_{V}}$, and thus a reduction in the entropy, due to the intermolecular attractive Van der Waals forces.  Equation \ref{eq:eqSTD_V_norm_regression_error_RF} also demonstrates that this reduction clearly decreases with both increasing reduced temperature $T_R$ and reduced volume $V_R$.  This parametric numerical simulation makes it clear that increasing the Van der Waals intermolecular force has a reducing impact on the entropy of a non-ideal fluid.  

\subsection{Stirling Cycle Simulation}

There are a few essential equations essential to a proper thermodynamic engine analysis, to characterize the properties of a thermodynamic heat engine.  First, specific work \emph{w} (J/kg), by definition, is the integral of the pressure over the volume, 
\begin{eqnarray}
\label{eq:eq_W_definition}
w&=&{\int}P{\cdot}{\delta}v,
\end{eqnarray}
where \emph{P} (Pa) represents the pressure, and ${\delta}v$ (m$^3$/kg) represent the change in specific volume.  This specific work \emph{w} (J/kg) is an important component for the first law of thermodynamics, represented in equation \ref{eq:eq_First_Law},
\begin{eqnarray}
\label{eq:eq_First_Law}
{\delta}u&=&q-w,
\end{eqnarray}
where ${\delta}u$ (J/kg) represents the change in specific internal energy, and \emph{q} (J/kg) represents the specific heat input or output.

A real-fluid Stirling cycle heat engine utilizing one mole of argon as the working fluid was simulated, ranging from a reduced specific volume $V_R$ of 1.5 to 30 (density $\rho$ ranging from 17.8333 kg/m$^3$ to 356.6666 kg/m$^3$), and a reduced temperature $T_R$ ranging from 1.2 to 2 (absolute temperatures \emph{T} ranging from 180.8244 K to 301.374 K).  The pressures \emph{P} (Pa) were calculated using the Peng-Robinson equation of state defined in equation \ref{eq:eqPengRobinson} \cite{PR1976,PitzerAcentric}, and the internal energy \emph{U} (J/mole) was estimated using equation \ref{eq:eqU_mytheory}, which was validated experimentally by the author in an earlier effort \cite{Marko_SciReports}.  All of these properties are tabulated in Table \ref{tb:tb_AR_thermo}.  

The work input and output \emph{W} (J/mole) of this cycle during compression (Stage 12) and expansion (Stage 34) was calculated using equation \ref{eq:eq_W_definition}, integrating the pressure solved with the Peng-Robinson equation of state over the change in volume.  The heat input and output \emph{Q} (J/mole) was solved by realizing the change in internal energy ${\delta}U$ (J/mole) and work \emph{W} (J/mole), utilizing the first-law of thermodynamics defined in equation \ref{eq:eq_First_Law}.  The change in entropy ${\delta}s_U$ (J/mole$\cdot$K) to the ambient universe was found with equation \ref{eq:eqSideal}, dividing the specific heat inputs and outputs \emph{Q} (J/mole) by the absolute temperature \emph{T} (K). In addition, the change in the normalized standard deviation of the molecular velocity ${\delta}{\bar{\sigma}}_V$ was calculated using equations \ref{eq:eqSTD_V_norm_regression_IG} and \ref{eq:eqSTD_V_norm_regression_error_RF}, and the correlation \emph{R} between the change in entropy to the universe ${\delta}s_U$ (J/mole$\cdot$K) and the change in the normalized standard deviation of the molecular velocity ${\delta}{\bar{\sigma}}_V$ was calculated for each change in stage.  All of these results are tabulated in Table \ref{tb:tb_AR_entropy}.  

When summing up the changes in entropy to the universe ${\delta}s_U$ (J/mole$\cdot$K) at each stage in Table \ref{tb:tb_AR_entropy}, the entirety of the cycle results in a net negative entropy of -1.4865 (J/mole$\cdot$K).  In addition, the efficiency (equation \ref{eq:eqEff_Thermo}) of this heat engine cycle $\eta$ is 0.43, exceeding the Carnot efficiency $\eta_C$ (equation \ref{eq:eqEffCarnot}) of 0.4.  
\begin{eqnarray}
{\eta}=&{\dfrac{3572}{8201+1612-1509}}=&0.43,\nonumber \\ \nonumber
{\eta_C}=&1-{\dfrac{1.2}{2}}=&0.4.
\end{eqnarray}
This negative entropy and efficiency that exceeds the Carnot efficiency is predicated on the accuracy of equation \ref{eq:eqU_mytheory} to best represent the change in internal energy of a non-ideal fluid, rather than the traditional approach to calculating the change in internal energy of a non-ideal fluid, described in equation \ref{eq:eqdU_ideal}; the author's prior work \cite{Marko_SciReports} demonstrated experimentally that equation \ref{eq:eqU_mytheory} is a closer match than equation \ref{eq:eqdU_ideal}.  Finally, all of the correlations \emph{R} tabulated in Table \ref{tb:tb_AR_entropy} exceed 0.99, making it abundantly clear that the normalized standard deviation of the molecular velocity is a good representation of the true entropy (i.e. disorder) of a fluid, both ideal gases and real fluids subjected to the intermolecular Van der Waals forces.  Equation \ref{eq:eqSTD_V_norm_regression_error_RF} shows a clear reduction in this disorder with reduced temperature and increased density (reduced specific volume), where the intermolecular Van der Waals forces are more significant.  An isothermal heat transfer process from an ideal gas to a real-fluid will result in a greater reduction in entropy from the ideal gas, versus the increase in entropy of the real fluid.  One can only conclude that the intermolecular attractive Van der Waals forces reduce the potential for disorder, reduce the overall entropy of the fluid, and can enable a macroscopic heat engine to enable a net reduction in entropy and efficiencies that exceed the Carnot efficiency, provided the working fluid is a real-fluid subjected to significant Van der Waals forces during the thermodynamic cycle.

\begin{table}[h]
\begin{center}
\begin{tabular}{ | c || c | c | c | c | c | c |}
\hline
Stage & $T_R$ & $V_R$ & U & P & T & Density \\
& & & (J/mole) & (MPa) & (K) & (kg/m$^3$) \\
\hline
1 & 1.2 & 30 & 2210 & 0.65289 & 180.8244 & 17.8333\\
2 & 1.2 & 1.5 & 1354 & 9.0051 & 180.8244 & 356.6666\\
3 & 2 & 1.5 & 2965 & 23.0727 & 301.374 & 356.6666\\	
4 & 2 & 30 & 3719 & 1.1157 & 301.374 & 17.8333\\
  \hline
\end{tabular}
\caption{Thermodynamic Properties of non-ideal Argon Stirling-cycle heat engine simulation.}
\label{tb:tb_AR_thermo}
\end{center}
\end{table}

\begin{table}[h]
\begin{center}
\begin{tabular}{ | c || c | c | c | c | c |}
\hline
Stage & W & Q & ${\delta}s_U$ & ${\delta}{\bar{\sigma}_{V}}$ & R\\
 & (J/mole)  & (J/mole) & (J/mole$\cdot$K) & & \\
\hline
12 & 3875 & -4732 & 26.1678 & 26.1678 & 0.9972\\
23 & 0 & 1612 & -6.8392 & -6.8392 & 0.99436\\
34 & -7447 & 8201 & -27.2115 & -27.2115 & 0.99736\\
41 & 0 & -1509 & 6.3964 & 6.3964 & 0.99342\\
  \hline
$\Sigma$ & -3572 & 3572 & -1.4865 & 0 & - \\
  \hline
\end{tabular}
\caption{Entropy and Energy Results of non-ideal Argon Stirling cycle heat engine simulation.  The value of ${\bar{\sigma}_{V}}$ is obtained with regression equations \ref{eq:eqSTD_V_norm_regression_IG} and \ref{eq:eqSTD_V_norm_regression_error_RF}.}
\label{tb:tb_AR_entropy}
\end{center}
\end{table}

\section{Design of the Engine}
\label{sec:Engine_Design}
\subsection{Engine Introduction}

A closed-loop, internally reversible, heat engine-heat pump apparatus was designed \cite{Marko_USPTO_HE_HP} and built to replicate the Carnot thermodynamic cycle (Appendix \ref{ap:appendix_Carnot_HE_cycle}) to generate motive power from a heat reservoir and a temperature differential to the ambient.  By using a real fluid in the heat-engine (in this design carbon dioxide CO$_2$ is used), the theoretical thermodynamic efficiency $\eta$ (equation \ref{eq:eqEff_Thermo}) is boosted beyond that of the ideal-gas Carnot efficiency $\eta_C$ (equation \ref{eq:eqEffCarnot}), yielding a greater net energy output.  This system is entirely actuated and controlled by pneumatic gas and controlled by valves, eliminating the need to use precision motors or brakes \cite{Marko_USPTO_HE_simple,Marko_USPTO_HP} to fix the piston position at different stages of the cycle.  

This engine contains four piston-cylinders (Parts 1-4) actuated by pneumatic ideal gas, which comes from a large high-pressure compressed gas cylinder (Part 5).  These piston-cylinders are hydraulic actuators, with fittings (Part 6) to connect the rods between Cylinder H1 (Part 1) and Cylinder C4 (Part 4), as well as a fitting (Part 7) between Cylinder C2 (Part 2) and Cylinder C3 (Part 3).  Cylinder H1 (Part 1) and Cylinder C4 (Part 4) are arranged such that the piston can move the full stroke of the cylinders; Cylinder C2 (Part 2) and Cylinder C3 (Part 3) are installed at half the distance apart so the pistons can only move half a stroke length.  

All of the four piston-cylinders, as well as the compressed gas cylinder, are connected via piping and controlled by valves (Parts 8-11).  In addition the pneumatic piping (Part 8) between Cylinder C4 (Part 4) and the compressed gas cylinder (Part 5) has direction control via check-valves, allowing for controlled flow in each direction.  Cylinder H1 (Part 1) has a smaller bore than the identically sized Cylinders C2, C3, and C4 (Parts 2-4).  These components are laid out in the following pneumatic configuration: the compressed gas cylinder (Part 5) is pneumatically connected to the non-piston-rod side of Cylinder C4 (Part 4) via Part 8; the piston-rod side of Cylinder C4 (Part 4) is connected to the piston-rod side of Cylinder H1 (Part 1) via Part 9; the non-piston-rod side side of Cylinder H1 (Part 1) is pneumatically connected to the non-piston-rod side of Cylinder C2 (Part 2) via Part 10; and the piston-rod side of Cylinder C2 (Part 2) is connected pneumatically to the non-piston-rod side of Cylinder C3 (Part 3) via Part 11.  Cylinder H1 (Part 1) is maintained at a hotter temperature throughout the cycle; all of the other components are maintained at the ambient room-temperature.  There are four separate unique working fluids; three of them (Parts 12-14) are ideal gases at ambient temperature in the ambient-temperature cylinders (Parts 2-5); as well as a real fluid (Part 15) with a critical temperature comparable to the hot temperature maintained for Cylinder H1 (Part 1).  

\begin{figure}
\centering
\includegraphics[width=\linewidth]{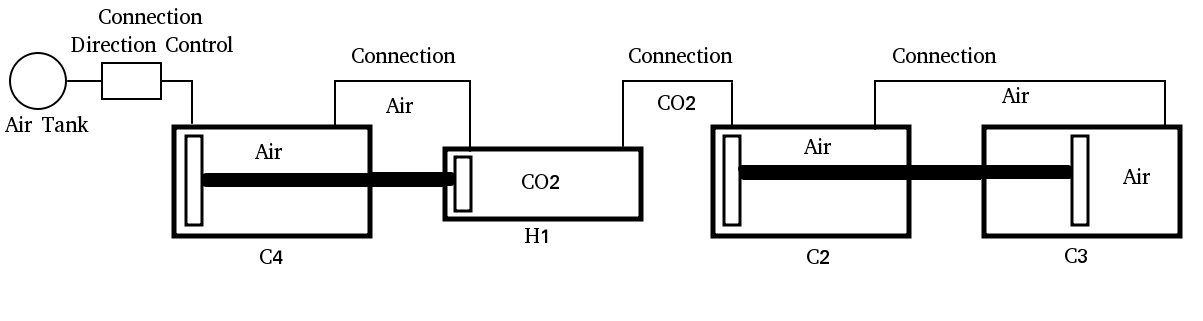}
\caption{The engine, with piston and working fluid positions arranged such that it is in Stage 1.  }
\label{fig:fig_Stage_1_5}
\end{figure}

The engine starts off at Stage 1 (Figure \ref{fig:fig_Stage_1_5}), where the ideal-gas working fluids (Parts 12-14) are all in the ambient temperature cylinders (Parts 2-5), and the real fluid (Part 15) is in the piston-rod side of the hot Cylinder H1 (Part 1).  Ideal-gas-working-fluid 1 (Part 12) is entirely in the large compressed gas cylinder (Part 5); ideal-gas-working-fluid 2 (Part 13) is entirely in the piston-rod side of Cylinder C4 (Part 4); and ideal-gas-working-fluid 3 (Part 14) entirely fills the piston-rod side of Cylinder C2 (Part 2) and the non-piston-rod side of Cylinder C3 (Part 3).  

\begin{figure}
\centering
\includegraphics[width=\linewidth]{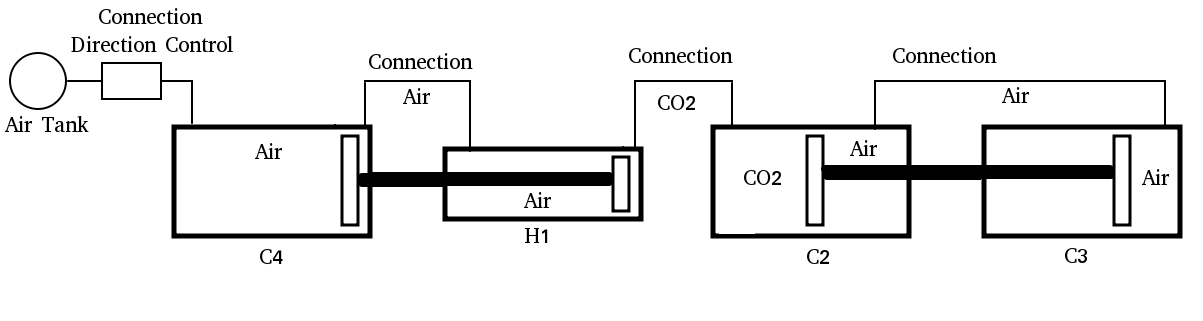}
\caption{The engine, with piston and working fluid positions arranged such that it is in Stage 2.  }
\label{fig:fig_Stage_2}
\end{figure}

To go to Stage 2 (Figure \ref{fig:fig_Stage_2}), ideal-gas-working-fluid 1 (Part 12) is allowed by the valve to flow from the compressed gas cylinder (Part 5) into Cylinder C4 (Part 4) via the direction controlled piping connection (Part 8).  This has the effect of forcing ideal-gas-working-fluid 2 (Part 13) from the larger Cylinder C4 (Part 4) into the smaller bore Cylinder H1 (Part 1) via the piping connection (Part 9); the movement of the piston in Cylinder C4 (Part 4) will actuate the piston in Cylinder H1 (Part 1), via the connecting rods (Part 6).  The compression of ideal-gas-working-fluid 2 (Part 13) from the larger Cylinder C4 (Part 4) to the smaller Cylinder H1 (Part 1) will compress this fluid, raising the temperature to exceed the hot temperature that the smaller cylinder Cylinder H1 (Part 1) is maintained at.  In addition, the movement of the piston in the smaller Cylinder H1 (Part 1) will force the carbon dioxide real-working-fluid (Part 15) from the smaller cylinder to the larger Cylinder C2 (Part 2) via the piping connection (Part 10); the expansion of the CO$_2$, as well as the work out to actuate the larger piston in Cylinder C2 (Part 2) will have effect of dropping the CO$_2$ temperature to that of the ambient temperature.  Finally, the movement of the piston in Cylinder C2 (Part 2) will both compress the ideal-gas-working-fluid 3 (Part 14), as well as actuate the movement of the piston in the equal-sized Cylinder C3 (Part 3), which is connected via the actuator rods (Part 7).  The movement of the connected pistons in Cylinder C2 (Part 2) and Cylinder C3 (Part 3) compresses the ideal-gas-working-fluid 3 (Part 14); which is contained both in the equally sized Cylinder C2 (Part 2) and Cylinder C3 (Part 3) cylinders, connected pneumatically by the valve-controlled piping (Part 11).  The pressure of the portion of ideal-gas-working-fluid 3 (Part 14) contained in the non-piston-rod side of Cylinder C3 (Part 3) will have a higher pressure than the pressure of the portion of ideal-gas-working-fluid 3 (Part 14) contained in the piston-rod side of Cylinder C2 (Part 2), due to the shorter effective volume at Stage 1, resulting in a greater relative compression for an equal piston actuation length.  

\begin{figure}
\centering
\includegraphics[width=\linewidth]{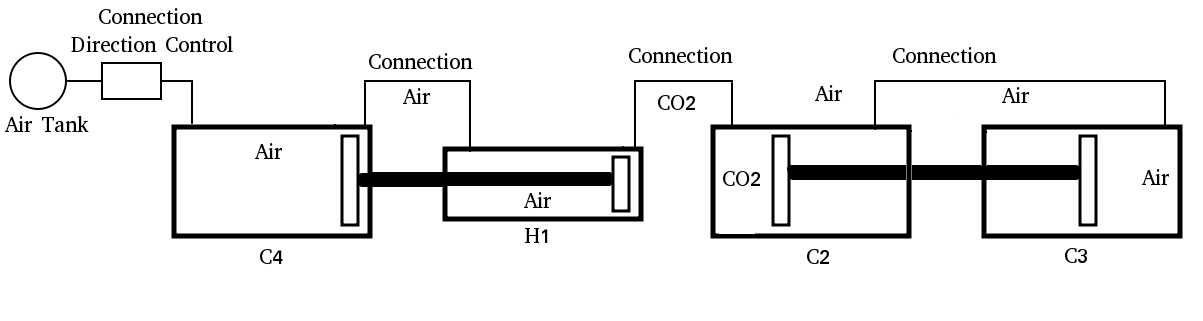}
\caption{The engine, with piston and working fluid positions arranged such that it is in Stage 3.  }
\label{fig:fig_Stage_3}
\end{figure}

To go to Stage 3 (Figure \ref{fig:fig_Stage_3}), the valve controlling the pneumatic piping (Part 11) between Cylinder C2 (Part 2) and Cylinder C3 (Part 3) is opened to allow flow of the ideal-gas-working-fluid 3 (Part 14) contained in Cylinder C3 (Part 3) to flow into Cylinder C2 (Part 2) so that the total pressure equalizes.  This will have the affect of compressing the real CO$_2$ working fluid (Part 15); it will remain in the larger Cylinder C2 (Part 2) because the piping (Part 10) connecting it to the smaller Cylinder H1 (Part 1) has a valve that is closed, and thus the carbon dioxide real-working-fluid (Part 15) remains at the ambient temperature at a greater pressure.  

\begin{figure}
\centering
\includegraphics[width=\linewidth]{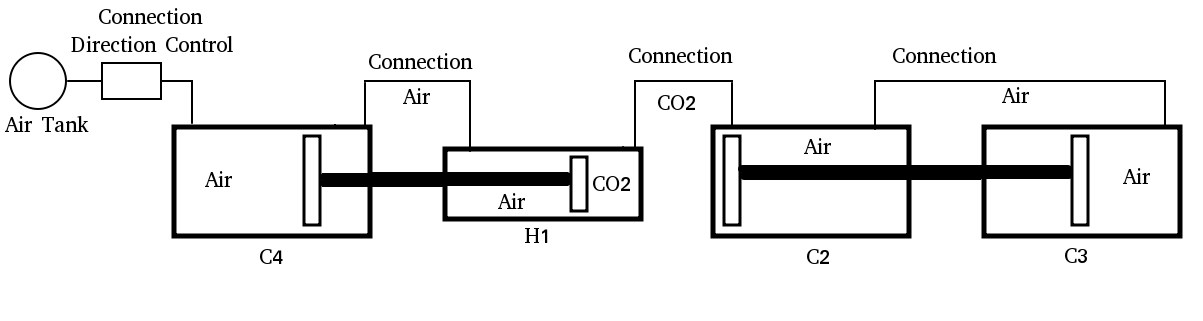}
\caption{The engine, with piston and working fluid positions arranged such that it is in Stage 4.  }
\label{fig:fig_Stage_4}
\end{figure}

To go to Stage 4 (Figure \ref{fig:fig_Stage_4}), the valve controlling the pneumatic piping (Part 10) between Cylinder H1 (Part 1) and Cylinder C2 (Part 2) is opened, allowing the CO$_2$ working fluid to flow into the smaller hot Cylinder H1 (Part 1), being compressed and thus having its temperature rise isentropically to match the hot cylinder’s temperature.  

To return to Stage 1 (Figure \ref{fig:fig_Stage_1_5}), the valve controlling the pneumatic piping (Part 8) between Cylinder C4 (Part 4) and the compressed gas cylinder (Part 5) is opened to allow flow back to the pressure vessel, allowing the ideal-gas-working-fluid 1 (Part 12) to flow back from Cylinder C4 (Part 4) into the compressed gas cylinder.  This will result in ideal-gas-working-fluid 2 (Part 13) flowing via the piping (Part 9) from Cylinder H1 (Part 1) back to Cylinder C4 (Part 4), causing the fluid to expand and cool, and absorb energy from the ambient temperature walls of Cylinder C4 (Part 4).  In addition, the carbon dioxide real-working-fluid will expand isothermally in Cylinder H1 (Part 1), resulting in an absorption of energy from the hot temperature source surrounding Cylinder H1 (Part 1).  At this point, only the carbon dioxide real-working-fluid (Part 15) shall remain in the smaller hot Cylinder H1 (Part 1), all of ideal-gas-working-fluid 1 (Part 12) has returned to the pressure vessel (Part 5), all of the ideal-gas-working-fluid 2 (Part 13) remains in Cylinder C4 (Part 4) at ambient temperature, and all of ideal-gas-working-fluid 3 (Part 14) remains within the larger bore Cylinders 2 (Part 2) and Cylinder 3 (Part 3) at an equal pressure.  When completed, all of the valves are closed, and the cycle can repeat itself.  

\begin{figure}[h]
\centering
\includegraphics[width=\linewidth]{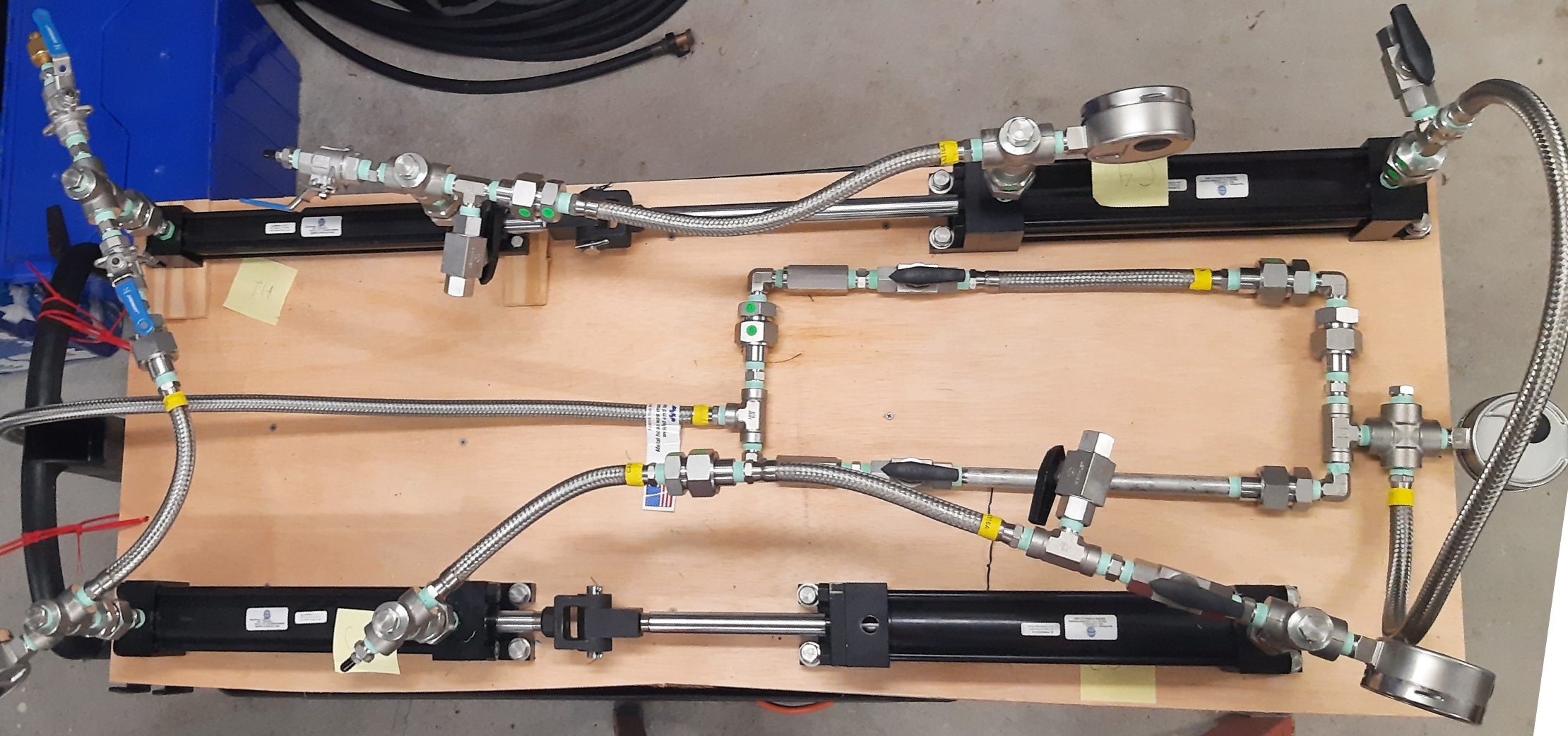}
\caption{Photograph of the Engine Apparatus}
\label{fig:fig_eng_photo1}
\end{figure}

A photographs of the built engine apparatus is included in Figure \ref{fig:fig_eng_photo1}.  

\clearpage

\subsection{Thermodynamic Cycle}

The heat engine utilizes the following thermodynamic cycle where a carbon dioxide real-working-fluid (Part 15) undergoes the following processes: 

\begin{itemize}
\item Stage 1-2: simultaneous compression and expansion (a net compression) from a hot gas in Cylinder H1 (Part 1) to a colder, higher-pressure gas in Cylinder C2 (Part 2), traveling via the pneumatic connection (Part 10) resulting in heat energy out to the ambient surrounding Cylinder C2 (Part 2).  
\item Stage 2-3: isothermal compression to a higher pressure in Cylinder C2 (Part 2), resulting in heat energy out to the ambient surrounding Cylinder C2 (Part 2).  
\item Stage 3-4: compression back to the original hot temperature while moving back from Cylinder C2 (Part 2) to Cylinder H1 (Part 1), traveling via the pneumatic connection (Part 10).  
\item Stage 4-1: isothermal expansion at the original hot temperature with Cylinder H1 (Part 1), resulting in heat energy being absorbed at the hot temperature source surrounding the smaller Cylinder H1 (Part 1). 
\end{itemize}
In addition, the ideal-gas-working-fluid 2 (Part 13) also goes through a thermodynamic process (with temperature changes) throughout the cycle: 
\begin{itemize}
\item Stage 1-2: compression from an ambient temperature gas in Cylinder C4 (Part 4) to a hotter, higher-pressure ideal gas in Cylinder H1 (Part 1), traveling via the pneumatic connection (Part 9), resulting in heat energy out to the hot temperature source surrounding Cylinder H1 (Part 1).  
\item Stage 2-3: no change.  
\item Stage 3-4: isothermal compression at the hot temperature, all within Cylinder H1 (Part 1), resulting in heat energy out to the hot temperature source surrounding Cylinder H1 (Part 1).  
\item Stage 4-1: expansion from the hot temperature in Cylinder H1 (Part 1), until it is fully returned to the larger Cylinder C4 (Part 4) and at the ambient temperature, resulting in heat absorption from the ambient surrounding Cylinder C4 (Part 4).  
\end{itemize}
Ideal-gas-working-fluid 1 (Part 12) and ideal-gas-working-fluid 3 (Part 14) remain consistently at the ambient temperature, and all compression and expansion are effectively isothermal, given the far greater specific heat capacity of the metallic cylinder walls.

\clearpage

\clearpage

\begin{figure}
\centering
\includegraphics[width=\linewidth]{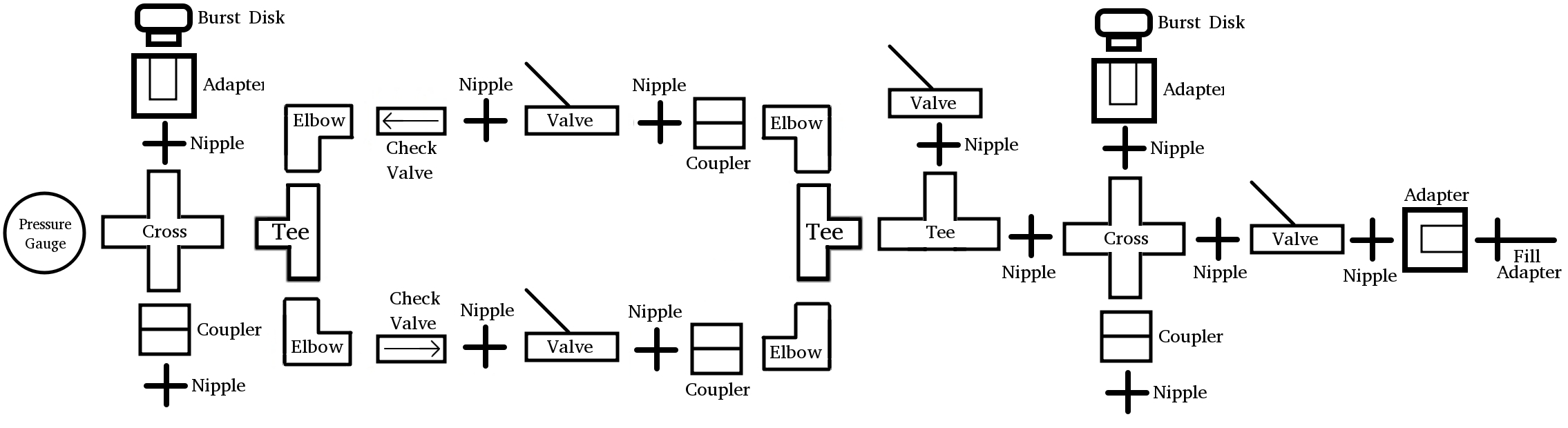}
\caption{The piping layout between the non-piston-rod side of C4 and the pressure vessel, for ideal-gas air, with direction control via check valves (Part 8).  }
\label{fig:fig_connection_with_dir}
\end{figure}

\begin{figure}
\centering
\includegraphics[width=\linewidth]{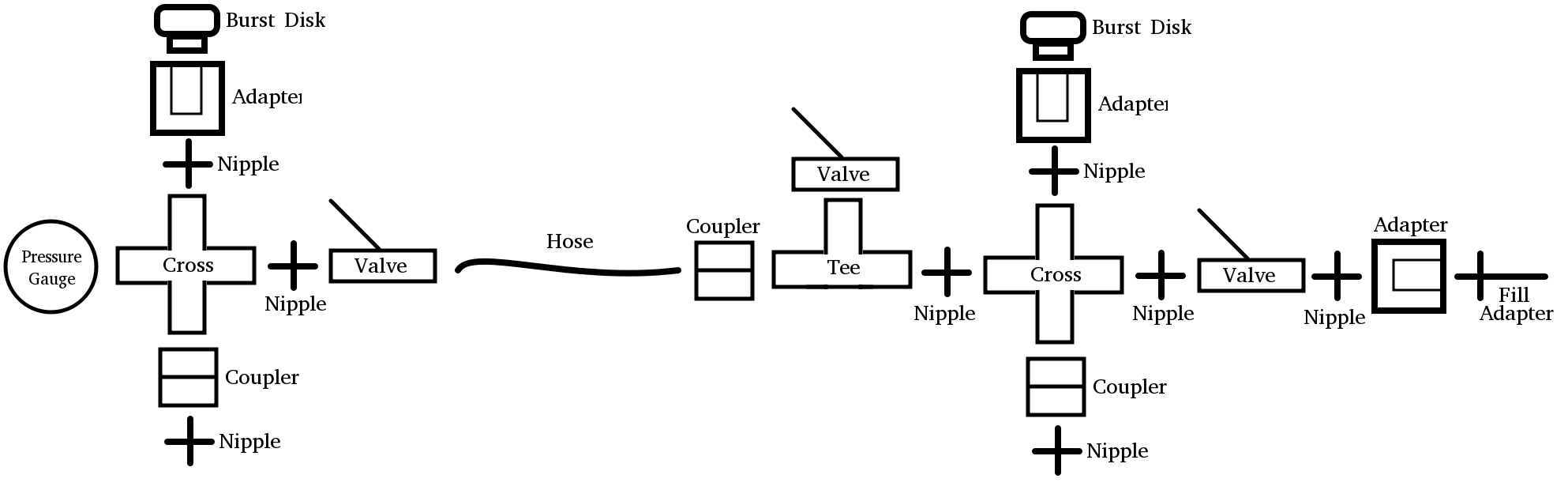}
\caption{The piping layout without direction control between the piston-rod sides of C4 and H1 (Part 9), between the non-piston-rod sides of H1 and C2 (Part 10), and between the piston-rod side C2 and the non-piston-rod side of C3 (Part 11).  }
\label{fig:fig_connection_no_dir}
\end{figure}

\clearpage

\section{Analysis of the Engine}
\label{sec:Engine_Analysis}
\subsection{Engine Equations}

Whenever an ideal gas is considered, a simplified equation for the specific internal energy \emph{u} (J/kg) is considered, based on equation \ref{eq:eqU_mytheory}, but removing the reduction in internal energy from the Van der Waals intermolecular attractive force, 
\begin{eqnarray}
\label{eq:eqU_Ideal_Gas}
{u}&=&{{C_V}{\cdot}{R_G}{\cdot}T},
\end{eqnarray}
where \emph{T} (K) represents the absolute temperature, $R_G$ (J/kg${\cdot}$K) represents the gas constant, and ${C_V}$ is equal to the number of degrees of freedom of the molecule plus one half (ex. monatomic fluids $C_V=1.5$, diatomic fluids $C_V=2.5$, etc).  When calculating the internal energy of an \emph{effective} ideal gas, both equations \ref{eq:eqU_mytheory} and \ref{eq:eqU_Ideal_Gas} match very closely.  For the sake of simplicity, equation \ref{eq:eqU_Ideal_Gas} will be used for calculating the internal energy of an ideal gas, which will be air in this engine analysis.  

The engine described in Section \ref{sec:Engine_Design} will be designed using both ambient air and carbon dioxide.  The thermodynamic properties will be obtained from NIST \cite{NIST_Webbook,CO2_NIST}, with the exception of the internal energy which will be derived from equation \ref{eq:eqU_mytheory} that was previously validated experimentally by the author \cite{Marko_SciReports}.  These properties include the critical temperature $T_C$ (K), the critical pressure $P_C$ (Pa), the critical density ${\rho}_C$ (kg/m$^3$), the degrees of freedom \emph{DOF} of the molecule, the molar mass \emph{MM} (kg/mole), the Pitzer's acentric factor $\omega$ defined in equation \ref{eq:eqPitzerAcentricFactor} \cite{PitzerAcentric}, the gas constant $R_G$ (J/kg$\cdot$K), the specific heat at a constant volume ${C_V}{\cdot}{R_G}$ (J/kg$\cdot$K), the specific heat at a constant pressure ${C_P}{\cdot}{R_G}$ (J/kg$\cdot$K), the specific heat ratio \emph{k}, 
\begin{eqnarray}
\label{eq:eqk_gas_ratio}
k&=&{\dfrac{C_P}{C_V}}={\dfrac{DOF+1.5}{DOF+0.5}},
\end{eqnarray}
the parameters for the Peng-Robinson equation of state defined in equation \ref{eq:eqPengRobinson} \cite{PR1976,PitzerAcentric}, and the calculated parameter \emph{a'} for the internal energy defined in equation \ref{eq:eqU_mytheory} \cite{Marko_SciReports}.  These properties are all listed in Table \ref{tb:tb_Eng_Thermo_Properties_Air_CO2}.  

The four cylinders will all have a stroke of 12 inches; Cylinder 3 and Cylinder 4 will be offset closer to each other by half the stroke length, an offset of 6 inches.  The bores of the 3 larger, ambient-temperature cylinders are all 4 inches; the bore of the smaller, heated cylinder is 2.5 inches.  The stand-alone pressure vessel will be 24 liters in volume.  The engine will be designed to operate between an absolute temperature range of $T_L$ = 293.15 K and $T_H$ = 322 K (68$^{\circ}$F and 120$^{\circ}$F).  The engine will contain 50 grams of carbon dioxide (\emph{RF}), 100 grams of air that is initially present in Cylinder 4 (\emph{IG 1}), 200 grams of air that is initially split between Cylinder 2 and Cylinder 3  (\emph{IG 2+3}), and 2.0 kilograms of air (\emph{IG0}) within the 24 liter pressure vessel \emph{C5}.  At an ambient temperature of $T_L$ = 293.15 K, the pressure of the pressure vessel \emph{C5} fluctuates between 844 psi and 1,108 psi.  These properties are all tabulated in Table \ref{tb:tb_Eng_Design_Properties_1}.  

All of the thermodynamic properties at the four stages are are tabulated in Table \ref{tb:tb_Eng_Thermo_Properties_1}.  This includes the absolute temperature \emph{T} (K), the total volume \emph{V} (cm$^3$), the pressure \emph{P} (MPa / psi), the density $\rho$ (kg/m$^3$), and the internal energy \emph{U} (J) calculated with equation \ref{eq:eqU_mytheory} for the carbon dioxide (\emph{RF}) and equation \ref{eq:eqU_Ideal_Gas} for the air (\emph{IG}).  The change in internal energy ${\delta}U$ (J) for each stage is calculated with equations \ref{eq:eqU_mytheory} and \ref{eq:eqU_Ideal_Gas}, and tabulated in Table \ref{tb:tb_Eng_Thermo_Properties_dU}; the ideal gas air has no change in internal energy ${\delta}U=0$ when there is no change in absolute temperature \emph{T} (K).  The calculated change in work ${\delta}w$ (J), solved with equation \ref{eq:eq_W_definition}, is tabulated in Table \ref{tb:tb_Eng_Thermo_Properties_dW}.  Finally, using the change in internal energy ${\delta}u$ (J) and the work ${\delta}w$ (J), the heat transfer \emph{q} (J) is found via the first-law of thermodynamics equation \ref{eq:eq_First_Law}; the results are tabulated in Table \ref{tb:tb_Eng_Thermo_Properties_Q}.  

\clearpage

\subsection{Engine Results}

\begin{table}[h]
\begin{center}
\begin{tabular}{ | c || c | c || c |}
\hline
 & Air & CO$_2$ & \\
\hline
T$_C$ (K) & 132.63 & 304.13 & Critical Temperature\\
P$_C$ (Pa) & 6234019 & 7377300 & Critical Pressure\\
$\rho_C$ (kg/m$^3$) & 231 & 468 & Critical Density\\
DOF & 2 & 3 & Degrees of Freedom\\
MM (kg/mole) & 0.02897 & 0.0440095 & Molar Mass\\
$\omega$ & 0.0362 & 0.228 &  Pitzer's Acentric Factor Equation \ref{eq:eqPitzerAcentricFactor} \\
\hline
R$_G$ (J/kg$\cdot$K) & 287.00 & 188.92 & Specific Gas Constant \\
C${_V}{\cdot}$R$_G$ (J/kg$\cdot$K) & 717.51 & 661.23 & Specific Heat, Constant Volume\\
C${_P}{\cdot}$R$_G$ (J/kg$\cdot$K) & 1,004.51 & 850.15 & Specific Heat, Constant Pressure\\
k & 1.40 & 1.29 & Specific Heat Ratio Equation \ref{eq:eqk_gas_ratio}\\
${\kappa}$ & 0.43 & 0.71 & Peng-Robinson Equation \ref{eq:eqPengRobinson}\\
A & 106.27 & 204.62 & Peng-Robinson Equation \ref{eq:eqPengRobinson}\\
B & 4.75E-04 & 6.06E-04 & Peng-Robinson Equation \ref{eq:eqPengRobinson}\\
a' & 249.86 & 728.48 & Internal Energy Equation \ref{eq:eqU_mytheory}\\
\hline
\end{tabular}
\caption{Properties of the Working Fluids}
\label{tb:tb_Eng_Thermo_Properties_Air_CO2}
\end{center}
\end{table}

\begin{table}[h]
\begin{center}
\begin{tabular}{ | c c || c |}
\hline
Vol H1 (m$^3$) & 1.93E-03 & 12” stroke, 2.5” bore\\
Vol C2 (m$^3$) & 4.94E-03 & 12” stroke, 4” bore\\
Vol C3 (m$^3$) & 2.47E-03 & 6” stroke, 4” bore\\
Vol C4 (m$^3$) & 4.94E-03 & 12” stroke, 4” bore\\
Vol C5 (m$^3$) & 0.024 & 2 12-liter tanks\\
\hline
T$_L$ (K)  & 293.15 & 68 $^{\circ}$F\\
T$_H$ (K)   & 322 & 119.93$^{\circ}$F\\
\hline
dL C3 (in) & 6 & \\
\hline
Mass RF (kg)   & 0.05 & \\
Mass IG-1 (kg)   & 0.1 & \\
Mass IG-2+IG-3 (kg)   & 0.2 & \\
Mass IG-2 (kg)   & 0.1333 & \\
Mass IG-3 (kg)   & 0.0667 & \\
Mass IG00 (kg)   & 2.0 & \\
\hline
a'$_{CO_2}$   & 728.48 & Equations \ref{eq:eqDeltaU_isothermal_mytheory} and \ref{eq:eqU_mytheory} for CO$_2$\\
a   & 204.62 & Peng-Robinson equation \ref{eq:eqPengRobinson} for CO$_2$\\
b   & 6.06${\cdot}{10^{-4}}$ & Peng-Robinson equation \ref{eq:eqPengRobinson} for CO$_2$\\
\hline
C5 P$_{High}$ (psi) & 1,018 & \\
C5 P$_{Low}$ (psi) & 844 & \\
\hline
\end{tabular}
\caption{Physical Properties of the Engine.  The subscript \emph{IG} refers to air as an ideal gas, and the subscript \emph{RF} refers to carbon dioxide as a real fluid.  }
\label{tb:tb_Eng_Design_Properties_1}
\end{center}
\end{table}

\clearpage

\begin{table}[h]
\begin{center}
\begin{tabular}{ | c | c || c | c | c | c | c | c | c |}
\hline
Stage & Fluid & T (K) & V (cm$^3$) & P (MPa) & P (psi) & $\rho$ (kg/m$^3$) & U (J)\\
\hline
1 & IG-1 & 293.15 & 4,940 & 1.70 & 247 & 20.23 & 21,033.69\\
1 & IG-2 & 293.15 & 4,940 & 2.27 & 330 & 26.98 & 28,044.91\\
1 & IG-3 & 293.15 & 2,470 & 2.27 & 330 & 26.98 & 14,022.46\\
1 & RF & 322 & 1,930 & 1.47 & 214 & 25.90 & 10,423.18\\
\hline
2 & IG-1 & 322 & 1,930 & 4.79 & 695 & 51.80 & 23,103.69\\
2 & IG-2 & 293.15 & 4,100 & 2.73 & 397 & 32.50 & 28,044.91\\
2 & IG-3 & 293.15 & 1,630 & 3.44 & 500 & 40.88 & 14,022.46\\
2 & RF & 293.15 & 840 & 2.72 & 396 & 59.51 & 9,168.19\\
\hline
3 & IG-1 & 322 & 1,930 & 4.79 & 695 & 51.80 & 23,103.69\\
3 & IG-2+3 & 293.15 & 5,850 & 2.88 & 418 & 22.79 & 42,067.37\\
3 & RF & 293.15 & 781 & 2.89 & 420 & 64.03 & 9,128.40\\
\hline
4 & IG-1 & 322 & 1,560 & 5.91 & 858 & 63.95 & 23,103.69\\
4 & IG-2 & 293.15 & 4,940 & 2.27 & 330 & 26.98 & 28,044.91\\
4 & IG-3 & 293.15 & 2,470 & 2.27 & 330 & 26.98 & 14,022.46\\
4 & RF & 322 & 367 & 5.89 & 856 & 136.31 & 9,473.80\\
\hline
\end{tabular}
\caption{Calculated thermodynamic properties at each stage of the engine cycle.}
\label{tb:tb_Eng_Thermo_Properties_1}
\end{center}
\end{table}

\begin{table}[h]
\begin{center}
\begin{tabular}{ | c || c | c | c | c |}
\hline
Stage  & IG-1 & IG-2+IG-3 & RF & IG-0\\
\hline
12 & 2070.00 & 0.00 & -1254.99 & 0.00\\
23 & 0.00 & 0.00 & -39.79 & 0.00\\
34 & 0.00 & 0.00 & 345.40 & 0.00\\
41 & -2070.00 & 0.00 & 949.38 & 0.00\\
\hline
\end{tabular}
\caption{The change in internal energy $\delta$U (J) for each change in stage during the engine cycle.}
\label{tb:tb_Eng_Thermo_Properties_dU}
\end{center}
\end{table}

\begin{table}[h]
\begin{center}
\begin{tabular}{ | c || c | c | c | c || c |}
\hline
Stage  & IG-1 & IG-2+IG-3 & RF & IG-0 & Total\\
\hline
12 & 9,771.83 & 4,500.62 & 2,288.59 & -5,917.07 & 10,643.97\\
23 & 0.00 & -166.48 & 166.48 & 0.00 & 0.00\\
34 & 1,961.82 & -4,017.93 & 1,817.79 & 0.00 & -238.32\\
41 & -12,858.78 & 0.00 & -5,757.54 & 5,917.07 & -12,699.25\\
\hline
Net & -1,125.13 & 316.21 & -1,484.68 & 0.00 & -2,293.60\\
\hline
\end{tabular}
\caption{The work input / output $\delta$W (J) for each change in stage during the engine cycle.}
\label{tb:tb_Eng_Thermo_Properties_dW}
\end{center}
\end{table}

\begin{table}[h]
\begin{center}
\begin{tabular}{ | c || c | c | c | c |}
\hline
Stage  & IG-1 & IG-2 + IG-3 & RF & IG-0\\
\hline
12 & \bf{\underline{-7,701.83}} & -4,500.62 & \bf{\underline{-3,543.58}} & 5,917.07\\
23 & \bf{0.00} & 166.48 & -206.27 & 0.00\\
34 & \bf{-1,961.82} & 4,017.93 & -1,472.39 & 0.00\\
41 & \bf{10,788.77} & 0.00 & \bf{6,706.92} & -5,917.07\\
\hline
\end{tabular}
\caption{The heat transfer \emph{Q} (J) in and out of the engine for each change in stage of the heat engine cycle.  The {\bf{Bold}} font represents hot heat transfer, the non-bold represents cold heat transfer to the ambient temperature.  The {\bf{\underline{Bold Underlined}}} numbers represent heat transfer that optimally is at the hot temperature, but conservatively is transferred to the cold ambient temperature. }
\label{tb:tb_Eng_Thermo_Properties_Q}
\end{center}
\end{table}

\clearpage

\subsection{Engine Efficiency}

According to Table \ref{tb:tb_Eng_Thermo_Properties_dW}, the total mechanical work output per cycle is 2,293.6 J; this work output overwhelming happens towards the end of the cycle stage 41.  This work can be connected to a flywheel or an electric energy generator, for whatever useful purpose is desired.  According to Table \ref{tb:tb_Eng_Thermo_Properties_Q}, the total hot heat input can range from 4,288.47 J to 15,533.88 J.  The reason for this discrepancy is that between Stage 12, the carbon dioxide real fluid \emph{RF} and the ideal gas air \emph{IG1} will release heat both within the heated smaller-bore cylinder \emph{H1} and within the ambient-temperature, larger-bore cylinders \emph{C2} and \emph{C4}.  Ideally, all of this heat will transfer out within the heated, smaller-bore cylinder to maximize the efficiency, but this is not a practical reality, therefore the hot heat input will be bounded between this optimistic and conservative estimation.  Using equation \ref{eq:eqEff_Thermo} for the thermodynamic efficiency, the conservative and optimistic thermodynamic efficiency will range between 14.77\% and 53.48\%, both of which are greater than the Carnot efficiency of 8.96\% calculated with equation \ref{eq:eqEffCarnot}.  

\begin{itemize}
\item Net W = 2,293.60
\item Net $Q_H$ = 15,533.88 (Conservative)
\item Net $Q_L$ = -13,249.28 (Conservative)
\item Net $Q_H$ = 4,288.47 (Optimistic)
\item Net $Q_L$ = -1,994.87 (Optimistic)
\item Efficiency $\eta$ = $\dfrac{2,293.60}{15,533.88}$ = 14.77\% (Conservative)
\item Efficiency $\eta$ = $\dfrac{2,293.60}{4,288.47}$ = 53.48\% (Optimistic)
\item Efficiency Carnot $\eta_C$ = 1 - $\dfrac{293.15}{322}$ = 8.96\%
\end{itemize}

\clearpage

\section{Conclusion}
\label{sec:Conclusion}
In conclusion, a practical, macroscopic-scale piston-cylinder engine was built and demonstrated \cite{Marko_USPTO_HE_HP}, utilizing a novel thermodynamic cycle.  This practical engine accomplished two things.  First, it represented a realistic approach to building a heat engine that closely resembled the Carnot heat engine cycle, utilizing an arrangement of valves and pneumatic air to achieve the (nearly) isothermal compression and expansion, as well as the (nearly) isentropic heating and cooling, of the working fluid.  This engine was built at the macroscopic scale, and it did not require any advanced or high-precision manufacturing, demonstrating its usefulness as a practical engine.  Second, by using a non-ideal working fluid (carbon dioxide CO$_2$), the engine was able to take advantage of the intermolecular attractive Van der Waals forces to boost the thermodynamic efficiency (equation \ref{eq:eqEff_Thermo}) of the heat engine, theoretically beyond the Carnot efficiency (equation \ref{eq:eqEffCarnot}).  As demonstrated experimentally by the author in a previous effort \cite{Marko_SciReports}, because the intermolecular Van der Waals forces are stronger at colder temperatures, these forces reduced the negative work input more than the positive work output, increasing the net total work output and thus boosting the overall efficiency.  This engine demonstrated this capability in a practical, macroscopic heat engine, and offers great opportunities for practical energy generation.  

\clearpage

\bibliographystyle{unsrt}
\bibliography{RefFile}

\clearpage
\newpage
\setcounter{section}{0}

\renewcommand{\thesection}{\Alph{section}}
\section{Appendix}

\subsection{Derivation of Carnot Efficiency}
\label{ap:appendix_Carnot}
The Carnot efficiency ${\eta}_C$ (equation \ref{eq:eqEffCarnot}) represents the theoretical maximum efficiency of a heat engine operating with an ideal-gas working fluid.  It is rooted in the second law of thermodynamics, which states definitively that heat, at a macroscopic level, can only flow from hot to cold; it can never flow from cold to hot.  Temperature, by definition, is the average root-mean-square speed of the molecules of a given thermal source or sink.  A hot thermal source fluid will have a higher average speed than a colder sink.  While there exists some infinitesimally small probability of faster-than-average molecules from the cold-sink transferring energy to the slower-than-average molecules in the hot source, in the aggregate, considering macroscopic sources and sinks operate with moles (6.022$\cdot$10$^{23}$) of particles, in time the hot thermal source will always transfer energy to the cold thermal sink.  

As a result of this, all heat transfer processes add to the net disorder in the universe; with every heat transfer process between a temperature differential, there is more and more net disorder in the universe.  Clausius' theorem (equation \ref{eq:eqClausius}) \cite{ClausiusOrig} was derived to represent the second law, stating that any internally reversible thermodynamic cycle must generate a positive entropy ${\delta}s{\geq}0$ to the surrounding universe, and defines a reversible process where the net entropy to the universe is equal to zero, where the change in specific entropy ${\delta}s$ (J/kg$\cdot$K) is defined in equation \ref{eq:eqSideal}, 
\begin{eqnarray}
{\delta}{s}&=&\frac{q}{T},\nonumber
\end{eqnarray}
where \emph{T} (K) is the absolute temperature, and \emph{q} (J/kg) represent the heat transferred per unit mass.  

The Carnot efficiency ${\eta}_C$ (equation \ref{eq:eqEffCarnot}) represents the thermodynamic efficiency where there is no net change in entropy, 
\begin{eqnarray}
{{\oint}{\frac{{\delta}q}{T}}}&=&0.\nonumber
\end{eqnarray}
A heat engine is any machine that takes a heat input at a hot temperature source $Q_{IN}$ (J), and has an output of both mechanical work ${W_{OUT}}$ (J) and leftover heat at the cold temperature sink $Q_{OUT}$ (J), where 
\begin{eqnarray}
\label{eq:eq_Q_W_HE}
{W_{OUT}}&=&{Q_{IN}}-{Q_{OUT}}.  
\end{eqnarray}
The efficiency of a heat engine is defined in equation \ref{eq:eqEff_Thermo} as the ratio work output $W_{OUT}$ (J) over the heat input $Q_{IN}$ (J), 
\begin{eqnarray}
{\eta}&=&{\frac{W_{OUT}}{Q_{IN}}}.  \nonumber
\end{eqnarray}
Plugging equation \ref{eq:eq_Q_W_HE} into equation \ref{eq:eqEff_Thermo}, 
\begin{eqnarray}
\label{eq:eq_HE_effic_all_Q}
{\eta}&=&{\frac{{Q_{IN}}-{Q_{OUT}}}{Q_{IN}}}.  \\ \nonumber
&=&1-{\frac{Q_{OUT}}{Q_{IN}}}.
\end{eqnarray}
If a heat engine is to have no net change in entropy ${\delta}s=0$, then the entropy change from the heat input must by equal to the entropy change from the heat output, 
\begin{eqnarray}
{{\delta}s_{OUT}}-{{\delta}s_{IN}}&=&0,\nonumber \\
{{\delta}s_{OUT}}&=&{{\delta}s_{IN}},\nonumber \\
\frac{Q_{OUT}}{T_L}&=&\frac{Q_{IN}}{T_H},\nonumber \\
\frac{Q_{OUT}}{Q_{IN}}&=&\frac{T_L}{T_H}.\nonumber
\end{eqnarray}
Since ${{Q_{OUT}}/{Q_{IN}}}={{T_L}/{T_H}}$, this value for the ratio of the heat output over the heat input can be plugged into the efficiency of a heat engine, to get the Carnot efficiency ${\eta}_C$ (equation \ref{eq:eqEffCarnot}), 
\begin{eqnarray}
{\eta_C}&=&1-{\frac{Q_{OUT}}{Q_{IN}}}=1-{\frac{T_L}{T_H}}.\nonumber
\end{eqnarray}

A heat pump is simply a reverse heat engine, where mechanical work $W_{IN}$ (J) is the input, and as a result the heat pump extracts heat from a cold source (refrigeration or air conditioning) and supplies it at a hotter source $Q_{OUT}$ (J). As the efficiency of a heat engine ${\eta}=W_{OUT}/{Q_{IN}}$ is the work output $W_{OUT}$ (J) over the hot heat input $Q_{IN}=Q_H$ (J), the Coefficient of Performance (COF) is simply the inverse efficiency, or the hot heat output versus the work input,
\begin{eqnarray}
\label{eq:eq_COP_HP}
{COP}&=&{\frac{1}{\eta}}={\frac{Q_{OUT}}{W_{IN}}},
\end{eqnarray}
and the Carnot efficiency limit for a heat pump utilizing an ideal gas working fluid is the inverse of the Carnot efficiency,
\begin{eqnarray}
\label{eq:eq_COP_HP_Carnot}
{COP_C}&=&{\frac{1}{\eta_C}}=1/\{{1-{\frac{T_L}{T_H}}}\}.
\end{eqnarray}

\begin{figure}[h]
\centering
\includegraphics[width=\linewidth]{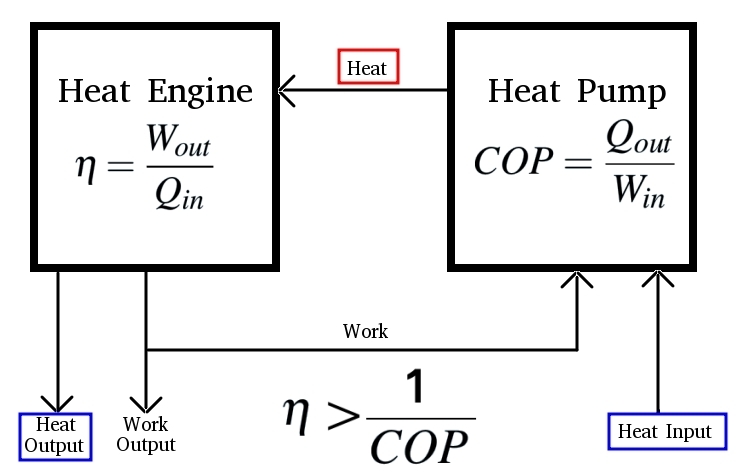}
\caption{Demonstration of significance of the Carnot efficiency ${\eta}_C$ (equation \ref{eq:eqEffCarnot}).}
\label{fig:fig_HE_HP}
\end{figure}

As a theoretical demonstration, one can envision a heat engine connected to a heat pump, as depicted in Figure \ref{fig:fig_HE_HP}. The heat output of the heat pump is the heat input of the heat engine, and the work output of the heat engine is the work input of the heat pump. If the efficiency of the heat engine were to exceed the inverse COP of the heat pump, then there would be an excess of work output of the heat engine, and the practical effect of these two machines is to extract heat from the ambient (cooling) and converting that energy to useful mechanical work.

Because of the second law of thermodynamics, which states that (macroscopically) heat must always flow from hot to cold, then the hot temperature output of the heat pump has to match or exceed the hot temperature input of the heat engine ($T_{H,HP}{\geq}{T_{H,HE}}$), assuming the cold temperature reservoir is the same ($T_{L,HP}=T_{L,HE}$).  If either the heat pump or the heat engine was less efficient than the Carnot limit, then an additional external work input or hot heat input will be needed to keep this device running, as the heat engine efficiency will be less than the inverse of the heat pump COP (${\eta_{HE}}<$ 1/COP$_{HP}$). If both the heat pump and heat engine were operating at the Carnot limit, and the heat pump’s hot-temperature output matched the heat engine’s hot-temperature input, then the heat engine's efficiency $\eta$ will match the inverse of the heat pump COP ($\eta_{HE}$ = 1/COP$_{HP}$), and the cycle can run indefinitely but there will never be an excess work output. If, however, the heat engine efficiency or the heat pump COP ever exceeded the Carnot limit, then ${\eta_{HE}}>$ 1/COP$_{HP}$ and the arrangement of Figure \ref{fig:fig_HE_HP} can exist. This is the reason for the significance of the Carnot efficiency and its relevance to the second law of thermodynamics.

\subsubsection{Carnot Heat Engine Cycle}
\label{ap:appendix_Carnot_HE_cycle}

Another demonstration of the Carnot efficiency ${\eta}_C$ (equation \ref{eq:eqEffCarnot}) is the Carnot heat engine cycle.  The Carnot heat engine is a theoretical thermodynamic cycle, utilizing an ideal gas, and designed to have no heat transfer across a significant temperature differential, and thus entirely reversible.  While all heat transfer, in practice, requires some temperature differential, the heat transfer from isothermal compression and expansion can effectively be treated as occurring at identical temperatures, and therefore entirely reversible, where the change in entropy to the universe ${\delta}s_U$ (J/K) is 0.  The Carnot cycle is one where, ideally, there is no generation of entropy \emph{s} (equation \ref{eq:eqSideal}), and the efficiency (equation \ref{eq:eqEff_Thermo}) matches the Carnot efficiency ${\eta}_C$ (equation \ref{eq:eqEffCarnot}).  

The stages of a Carnot cycle are as follows, 
\begin{itemize}
\item {\bf{Stage 12}}: Isothermal compression at the cold temperature $T_L$ (K) sink.
\item {\bf{Stage 23}}: Isentropic compression from the cold temperature $T_L$ (K) to the hot temperature $T_H$ (K).
\item {\bf{Stage 34}}: Isothermal expansion at the hot temperature $T_H$ (K) source.
\item {\bf{Stage 41}}: Isentropic expansion from the hot temperature $T_H$ (K) to the cold temperature $T_L$ (K).
\end{itemize}

During isentropic compression or expansion of an ideal gas, the work \emph{w}  (J/kg) will simply be equal to the change in internal energy of an ideal gas (equation \ref{eq:eqU_Ideal_Gas}), and the heat transfer \emph{Q} (J/kg) will be zero, by the definition of isentropic.  Assuming the working fluid is an ideal gas, $w_{23}=-w_{41}$, and $q_{23}=q_{41}=0$.  

During isothermal compression and expansion, the change in temperature is constant, thus the change in internal energy is constant (equation \ref{eq:eqU_Ideal_Gas}), and thus the heat input / output must be equal to the work output / input.  Plugging in the ideal gas pressure (equation \ref{eq:eqIdealGas}) into the equation for work (equation \ref{eq:eq_W_definition}), 
\begin{eqnarray}
w&=&{\int}P{\cdot}{\delta}v, \nonumber \\
&=&{\int}({\dfrac{{R_G}{\cdot}{T}}{{v}}}){\cdot}{\delta}v, \nonumber \\
&=&{{R_G}{\cdot}{T}}{\cdot}{\int}{\dfrac{{\delta}v}{{v}}}, \nonumber \\
&=&{{R_G}{\cdot}{T}}{\cdot}{log(\dfrac{v_2}{v_1})}, \nonumber\\
q&=&{{R_G}{\cdot}{T}}{\cdot}{log(\dfrac{v_1}{v_2})}. \nonumber
\end{eqnarray}
As these are ideal gases, the volume ratio of isentropic compression and expansion for an equal temperature must match \cite{2}, 
\begin{eqnarray}
{\dfrac{T_H}{T_L}}&=&{(\dfrac{v_2}{v_3})}^{k-1}={(\dfrac{v_1}{v_4})}^{k-1}, \nonumber 
\end{eqnarray}
where \emph{k} is the dimensionless specific heat ratio defined in equation \ref{eq:eqk_gas_ratio}, and therefore, 
\begin{eqnarray}
{\dfrac{v_2}{v_1}}&=&{\dfrac{v_3}{v_4}}. \nonumber 
\end{eqnarray}
The heat input $q_{34}$ (J) and heat output $q_{12}$ (J) is therefore,
\begin{eqnarray}
q_{in}=q_{34}&=&{{R_G}{\cdot}{T_H}}{\cdot}{log(\dfrac{v_3}{v_4})}, \nonumber \\
&=&{{R_G}{\cdot}{T_H}}{\cdot}{log(\dfrac{v_2}{v_1})}.\nonumber \\
q_{out}=q_{12}&=&{{R_G}{\cdot}{T_L}}{\cdot}{log(\dfrac{v_2}{v_1})}. \nonumber 
\end{eqnarray}
Plugging the known heat input $q_{in}$ (J) and the heat output $q_{out}$ (J), the efficiency $\eta$ can be found with equation \ref{eq:eq_HE_effic_all_Q}.  
\begin{eqnarray}
{\eta}&=&1-{\frac{q_{out}}{q_{in}}},\nonumber \\
&=&1-{\dfrac{{{R_G}{\cdot}{T_L}}{\cdot}{log(\dfrac{v_2}{v_1})}}{{{R_G}{\cdot}{T_H}}{\cdot}{log(\dfrac{v_2}{v_1})}}},\nonumber \\
&=&1-{\dfrac{T_L}{T_H}},\nonumber
\end{eqnarray}
which matches the Carnot efficiency $\eta_C$ defined in equation \ref{eq:eqEffCarnot}.  

\clearpage

\subsection{Engine Specifications}
\label{ap:appendix_engine_specifications}
\subsubsection{Part List}

\begin{enumerate}
\item Cylinder H1: a piston-cylinder hydraulic actuator, 2.5” diameter, 12” stroke
\item Cylinder C2: a piston-cylinder hydraulic actuator, 4” diameter, 12” stroke 
\item Cylinder C3: a piston-cylinder hydraulic actuator, 4” diameter, 12” stroke 
\item Cylinder C4: a piston-cylinder hydraulic actuator, 4” diameter, 12” stroke 
\item A network of compressed gas cylinders acting as a pressure vessel
\item Connection of hydraulic actuator piston rods for Part (1) and Part (4)
\item Connection of hydraulic actuator piston rods for Part (2) and Part (3)
\item Pneumatic Connection between Part (4) and Part (5), ideal gas with direction control
\item Pneumatic Connection between Part (4) and Part (1), ideal gas without direction control
\item Pneumatic Connection between Part (1) and Part (2), non-ideal CO$_2$ without direction control
\item Pneumatic Connection between Part (2) and Part (3), ideal gas without direction control
\item Ideal gas (air or nitrogen) working fluid 1 
\item Ideal gas (air or nitrogen) working fluid 2 
\item Ideal gas (air or nitrogen) working fluid 3
\item Real working fluid (CO$_2$)
\end{enumerate}

\clearpage

\subsubsection{Detailed Layout of Parts (8-11) Pneumatic Connections}

\noindent {\bf{(Part 8) Pneumatic Connection between Part (4) and Part (5), ideal-gas with direction control} - Figure \ref{fig:fig_connection_with_dir}}
\begin{itemize}
\item Connection Cross, F-F-F-F, 1/2" NPT 
\begin{itemize}
\item Pressure Gauge: M, 1/2" NPT, up to 5000 psi
\item Union Connector (to Cylinder C4), F-F, 1/2" NPT 
\begin{itemize}
\item Nipple, Quantity 2, M-M, 1/2" NPT 
\end{itemize}
\item Burst disk	
\begin{itemize}
\item Reducer, 1/2"-1/4" M-M NPT
\item Adapter, 1/4"-NPT – 3/8-24 UNF
\item Burst disk, 1800 psi – 3/8-24 UNF
\end{itemize}
\end{itemize}
\item Tee, F-M-F, 1/2" NPT 
\begin{itemize}
\item Elbow, Quantity 2, M-M, 1/2" NPT 
\item Union Connector, Quantity 2, F-F, 1/2" NPT 
\item Hose + pipe, Length 12", M-M, 1/2" NPT 
\begin{itemize}
\item One flexible hose
\item One rigid pipe
\end{itemize}
\item Ball valve, Quantity 2, F-F, 1/2" NPT 
\item Nipple, Quantity 2, M-M, 1/2" NPT
\item Check Valve, Quantity 2, F-F, 1/2" NPT
\begin{itemize}
\item Directions in reverse
\end{itemize}
\item Elbow, Quantity 2, M-M, 1/2" NPT
\end{itemize}
\item Tee, F-F-F, 1/2" NPT
\item Tee, M-M-M, 1/2" NPT
\begin{itemize}
\item Ball valve, F-F, 1/2" NPT (to bleed system)
\end{itemize}
\item Union Connector, F-F, 1/2" NPT
\item Reducer, 1/2"-1/4" M-M
\item Adapter to DIN, 1/4” F, NPT – DIN 0.750-14 NPSM
\end{itemize}

\noindent {\bf{(Part 9) Pneumatic Connection between Part (4) and Part (1), ideal-gas without direction control} - Figure \ref{fig:fig_connection_no_dir}}

\begin{itemize}
\item Connection Cross: F-F-F-F, 1/2" NPT
\begin{itemize}
\item Pressure Gauge: M, 1/2" NPT, up to 5000 psi
\item Union Connector (to Cylinder C4 / H1), F-F, 1/2" NPT
\begin{itemize}
\item Nipple, Quantity 2, M-M, 1/2" NPT
\end{itemize}
\item Burst disk	
\begin{itemize}
\item Reducer, 1/2"-1/4" M-M NPT
\item Adapter, 1/4"-NPT – 3/8-24 UNF
\item Burst disk, 1800 psi – 3/8-24 UNF
\end{itemize}
\end{itemize}
\item Nipple, M-M, 1/2" NPT 
\item Ball valve, F-F, 1/2" NPT
\item Hose, Length 12”, M-M, 1/2" NPT
\item Union Connector, F-F, 1/2" NPT
\item Tee, M-M-M, 1/2" NPT
\begin{itemize}
\item Ball valve, F-F, 1/2" NPT (to bleed system)
\end{itemize}
\item Connection Cross: F-F-F-F, 1/2" NPT
\begin{itemize}
\item Union Connector (to Cylinder H1 / C4), F-F, 1/2" NPT
\begin{itemize}
\item Nipple, Quantity 2, M-M, 1/2" NPT
\end{itemize}
\item Burst disk
\begin{itemize}
\item Reducer, 1/2"-1/4” M-M NPT
\item Adapter, 1/4"-NPT – 3/8-24 UNF
\item Burst disk, 1800 psi – 3/8-24 UNF
\end{itemize}
\item Air Fill
\begin{itemize}
\item Nipple, M-M, 1/2" NPT
\item Ball Valve, F-F, 1/2" NPT
\item Reducer, NPT, 1/2"-M to 1/8"-F
\item Air fill adapter, 1/8"-F NPT – 8-mm quick connect
\end{itemize}
\end{itemize}
\end{itemize}

\noindent {\bf{(Part 10) Pneumatic Connection between Part (1) and Part (2), Carbon Dioxide (CO$_2$) without direction control} - Figure \ref{fig:fig_connection_no_dir}}

\begin{itemize}
\item Connection Cross: F-F-F-F, 1/2" NPT
\begin{itemize}
\item Pressure Gauge: M, 1/2" NPT, up to 5000 psi
\item Union Connector (to Cylinder C2 / H1), F-F, 1/2" NPT
\begin{itemize}
\item Nipple, Quantity 2, M-M, 1/2" NPT
\end{itemize}
\item Burst disk	
\begin{itemize}
\item Reducer, 1/2"-1/4" M-M NPT
\item Adapter, 1/4"-NPT – 3/8-24 UNF
\item Burst disk, 1800 psi – 3/8-24 UNF
\end{itemize}
\end{itemize}
\item Nipple, M-M, 1/2" NPT 
\item Ball valve, F-F, 1/2" NPT
\item Hose, Length 12”, M-M, 1/2" NPT
\item Union Connector, F-F, 1/2" NPT
\item Tee, M-M-M, 1/2" NPT
\begin{itemize}
\item Ball valve, F-F, 1/2" NPT (to bleed system)
\end{itemize}
\item Connection Cross: F-F-F-F, 1/2" NPT
\begin{itemize}
\item Union Connector (to Cylinder H1 / C2), F-F, 1/2" NPT
\begin{itemize}
\item Nipple, Quantity 2, M-M, 1/2" NPT
\end{itemize}
\item Burst disk
\begin{itemize}
\item Reducer, 1/2"-1/4” M-M NPT
\item Adapter, 1/4"-NPT – 3/8-24 UNF
\item Burst disk, 1800 psi – 3/8-24 UNF
\end{itemize}
\item Air Fill
\begin{itemize}
\item Nipple, M-M, 1/2" NPT
\item Ball Valve, F-F, 1/2" NPT
\item Reducer, NPT, 1/2"-M to 1/4"-F
\item CO$_2$ fill adapter, 1/4" NPT – CGA-320
\end{itemize}
\end{itemize}
\end{itemize}

\noindent {\bf{(Part 11) Pneumatic Connection between Part (2) and Part (3), air without direction control} - Figure \ref{fig:fig_connection_no_dir}}

\begin{itemize}
\item Connection Cross: F-F-F-F, 1/2" NPT
\begin{itemize}
\item Pressure Gauge: M, 1/2" NPT, up to 5000 psi
\item Union Connector (to Cylinder C2 / C3), F-F, 1/2" NPT
\begin{itemize}
\item Nipple, Quantity 2, M-M, 1/2" NPT
\end{itemize}
\item Burst disk	
\begin{itemize}
\item Reducer, 1/2"-1/4" M-M NPT
\item Adapter, 1/4"-NPT – 3/8-24 UNF
\item Burst disk, 1800 psi – 3/8-24 UNF
\end{itemize}
\end{itemize}
\item Nipple, M-M, 1/2" NPT 
\item Ball valve, F-F, 1/2" NPT
\item Hose, Length 12”, M-M, 1/2" NPT
\item Union Connector, F-F, 1/2" NPT
\item Tee, M-M-M, 1/2" NPT
\begin{itemize}
\item Ball valve, F-F, 1/2" NPT (to bleed system)
\end{itemize}
\item Connection Cross: F-F-F-F, 1/2" NPT
\begin{itemize}
\item Union Connector (to Cylinder C3 / C2), F-F, 1/2" NPT
\begin{itemize}
\item Nipple, Quantity 2, M-M, 1/2" NPT
\end{itemize}
\item Burst disk
\begin{itemize}
\item Reducer, 1/2"-1/4” M-M NPT
\item Adapter, 1/4"-NPT – 3/8-24 UNF
\item Burst disk, 1800 psi – 3/8-24 UNF
\end{itemize}
\item Air Fill
\begin{itemize}
\item Nipple, M-M, 1/2" NPT
\item Ball Valve, F-F, 1/2" NPT
\item Reducer, NPT, 1/2"-M to 1/8"-F
\item Air fill adapter, 1/8"-F NPT – 8-mm quick connect
\end{itemize}
\end{itemize}
\end{itemize}

\clearpage

\clearpage
\subsection{Fortran Code}
\label{ap:appendix_fortran_code}

\begin{verbatim}

      program MakeInput
      implicit none

      real Vr,Tr,Vr_Fct(200),Tr_Fct(200)
      real U,S_entropy,P_kinetic,P_PR
      double precision pi,Kb,Av
      real t1,t2,Rat(100),outputdat(22)
      integer fooint,ct,ctx,ii,jj,kk,ppnum
      character(len=17) filenameSV


      call CPU_Time(t1)

      pi=3.1415926535897932384626
      Kb=1.38064852e-23  ! Boltzman's Constant
      Av=6.02214086e23   ! Avogadro's Number


c ---- Run a parametric series, first four separate heating simulations, 
c ---- then four stages of the high-pressure Stirling cycle



      ct=0
      do ii=1,20
        Tr=1.0*exp((ii-1)*0.10)
        do jj=1,10
          ct=ct+1
          Vr=1.0*exp((jj-1)*0.25)
          Tr_Fct(ct)=Tr
          Vr_Fct(ct)=Vr
        enddo
      enddo

      ctx=200
      open(unit=1000,file='param_output_data_study_8sept2020.txt')
      ppnum=0

      do ii=1,200
        ppnum=ppnum+1
        Tr=Tr_Fct(ii)
        Vr=Vr_Fct(ii)
        call ThermoCalc(Vr,Tr,ppnum,outputdat)
        call CPU_Time(t2)

        if (ppnum<10) then
          write(filenameSV,'("Save_00",I1,".txt")')ppnum
        elseif (ppnum<100) then
          write(filenameSV,'("Save_0",I2,".txt")')ppnum
        elseif (ppnum<1000) then
          write(filenameSV,'("Save_",I3,".txt")')ppnum
        endif
        open(unit=ppnum,file=filenameSV)
        write(ppnum,*) ppnum,Vr,Tr,outputdat,'t (s) = ',(t2-t1)
        close(ppnum)


        write(1000,*) ppnum,Vr,Tr,outputdat,'t (s) = ',(t2-t1)
        print *,ppnum,'/',ctx,'t (s) = ',t2
      enddo


      close(1000)

      end program


c ------------------------------- Analysis Subroutine ----------------------------

      subroutine ThermoCalc(Vr,Tr,ppnum,outputdat)

      real, intent(in) :: Vr,Tr
      integer, intent(in) :: ppnum
      real, intent(out) :: outputdat(22)


      integer ctsplitrng,ctsplit,fooint
      
      real U,S_entropy,P_kinetic,P_PR
      real t1,t2
      double precision  MM,Pc,Tc,Vc,ecc,V,T,Rg,a,b,R,AreaS,minVr
      double precision  kappa,a_PR,xx(3),Vx(3),Vx0(3),RMS3(3)
      double precision  Coeff
      double precision  dP_VWD, F_VDW_m,phi,theta,Vel,dP_VDW,V_rms_0,V_avg_0
      double precision drX,Vel_travel(3),Vel_xx
      double precision VelF,Vrat(3),Xrat(3)
      double precision VXdot,dP_Pauli,P_IG,U_KE,U_PE
      double precision Read6(6),Avg6(6),StDev6(6),V_rms_calc,V_avg_calc

      character (len=200) output
      integer ii,ii0,jj,jj0,kk,ct,dir(3)
      integer dx0, Nx, Ny, Np, Total_CT
      double precision bp,aa
      double precision pi,Kb,Av,m_m,dt,Fx(3),fooV(2)
      double precision, allocatable :: phi_fct(:),theta_fct(:),rrX(:,:)
      double precision, allocatable :: VXrat(:),F_dat(:),VelFdat(:)
      double precision, allocatable :: Vel_travelDat(:),Vel_fct(:)
      double precision, allocatable :: Xstore(:,:),Vstore(:,:)
      double precision, allocatable :: Fx_Stored(:,:)
      double precision, allocatable :: foocrap(:),randnum(:)
      integer, allocatable :: Tct(:),Total_Ct_Fct(:)

      character(len=15) filenameSV

      call CPU_Time(t1)

c ------------------------------- Make Source -------------------------------


      pi=3.1415926535897932384626
      Kb=1.38064852e-23 
      Av=6.02214086e23

      dx0=300   ! Estimated time steps per bounce
      Nx=91   ! Number of steps in each degree (square it in theta and phi)
      Ny=101   ! Number of steps at each degree increment, varying speed randomly
      ctsplit=500  ! Number of files to save all data (avoid limit of memory)


c --- Argon
      MM=.0399  ! Molar Mass (kg/mole)
      Pc=4.863e6  ! Critical Pressure (Pa)
      Tc=150.687  ! Critical Temperature (K)
      Vc=1./535    ! Critical Volume (m^3/kg)
      ecc=0       ! Eccentricity factor



c__________________________________________________

      Np=(Nx**2)*Ny  ! Total number of molecules bouncing in the spherical container
      
      if ((mod(Np,ctsplit))==0) then
        ctsplitrng=(Np/ctsplit)
      else
        ctsplitrng=(Np/ctsplit)+1
      endif
      print *,ctsplitrng

      ALLOCATE(rrX(Np,3))
      ALLOCATE(Tct(Np))
      ALLOCATE(VXrat(Np))
      ALLOCATE(F_dat(Np))
      ALLOCATE(VelFdat(Np))
      ALLOCATE(Vel_travelDat(Np))
      ALLOCATE(Vel_fct(Np))
      ALLOCATE(Xstore((dx0*10),3))
      ALLOCATE(Vstore((dx0*10),3))
      ALLOCATE(Fx_Stored((dx0*10),3))
      ALLOCATE(randnum(Ny))
      ALLOCATE(Total_Ct_Fct(ctsplit))

      ALLOCATE(foocrap(Np))

      V=Vr*Vc*MM  ! Actual volume (m^3)
      T=Tr*Tc     ! Actual temperature (K)

      m_m=MM/Av   ! Mass of Argon molecule
      Rg=Av*Kb/MM ! Gas Constant

c --- Peng-Robinson Equation of State coefficients
      a=0.45724*(Rg**2)*(Tc**2)/Pc
      b=0.07780*Rg*Tc/Pc
      
c --- Ensure the volume is not excessively small
      minVr=1./100
      if (Vr<((1+minVr)*b/Vc)) then
            V=((1+minVr)*b/Vc)*Vc*MM
      endif

c --- Coefficient for change in internal energy      
      bp=(2.**(1./3))-1.
      aa=(1/(9*bp))*(Rg**2)*(Tc**2.5)/Pc
     

c --- R = Radius of Sphere; Rb = equivalent radius of sphere (Pauli Exclusion)

      R=(V*(3./(4*pi)))**(1./3)
      Rb=((V-(b*MM))*(3./(4*pi)))**(1./3)
      AreaS=4*pi*(R**2) ! Surface Area of sphere

      V_rms_0=sqrt(3*Kb*T/m_m) ! RMS velocity of molecule
      V_avg_0=V_rms_0*(sqrt(8/(3*pi))) ! Average velocity of molecule

c --- Determine pressure from Peng-Robinson Equation of State
      kappa=0.37464+(1.54226*ecc)-(0.26992*(ecc**2))
      a_PR=(1.+(kappa*(1-(sqrt(T/Tc)))))**2
      P_PR=((Rg*T)/((V/MM)-b))
      P_PR=P_PR-((a_PR*a)/((((V/MM)**2)+(2*b*(V/MM))-(b**2))))


      dt=(2*R/V_avg_0)/dx0 ! Time step (s)


c --- Set function of angles in X-Y-Z coordinates
      ct=0
      do jj=1,Nx
            phi=(pi/2)*((jj-1.)/(Nx-1.))
            do ii=1,Nx
                  ct=ct+1
                  theta=(pi)*((ii-1.)/(Nx-1.))
                  xx(1)=(sin(theta))*(cos(phi))
                  xx(2)=(sin(theta))*(sin(phi))
                  xx(3)=cos(theta)


                  do kk=1,Ny
                        rrX(((ct-1)*Ny)+kk,1)=xx(1)
                        rrX(((ct-1)*Ny)+kk,2)=xx(2)
                        rrX(((ct-1)*Ny)+kk,3)=xx(3)
                  enddo
            enddo
      enddo

c --- Call velocity spread function (if Ny>1)
      call make_rand_fct(Ny,randnum)

c --- Confirm ratio of V_rms_calc / V_avg_calc = (pi*4/3)^(1/3)
      V_rms_calc=0
      V_avg_calc=0
      do ii=1,Ny
        V_avg_calc=V_avg_calc+(randnum(ii)*V_avg_0/Ny)
        V_rms_calc=V_rms_calc+(((randnum(ii)*V_avg_0)**2)/Ny)
      enddo
      V_rms_calc=sqrt(V_rms_calc)


c --- Set velocity function
      do ii=1,(Nx**2)
        do jj=1,Ny
          ii0=((ii-1)*Ny)+jj
          if (Ny==1) then
            Vel_fct(ii0)=V_rms_0
          else
c            Vel_fct(ii0)=V_rms_0
            Vel_fct(ii0)=(randnum(jj))*V_avg_0
          endif
        enddo
      enddo 


c ------------------------------- Make Source -------------------------------


      ii=0
      Total_CT=0
      Total_CT0=0

c --- Actually run the simulation, and saving X-Y-Z data of molecule until
c --- it reaches the opposing surface of the spherical container

      do ii0=1,ctsplit

        if ((ii+ctsplitrng)>Np) then
          fooint=Np-ii
        else
          fooint=ctsplitrng
        endif

        if (ii0<10) then
          write(filenameSV,'("SaveXV3_00",I1,".txt")')ii0
        elseif (ii0<100) then
          write(filenameSV,'("SaveXV3_0",I2,".txt")')ii0
        elseif (ii0<1000) then
          write(filenameSV,'("SaveXV3_",I3,".txt")')ii0
        endif
        open(unit=ii0,file=filenameSV)
        do jj0=1,fooint
          ii=ii+1
      
          call Get_dP_VDW(Vel_fct(ii),Vr,dP_VDW)
          F_VDW_m=dP_VDW*AreaS/Av
          Vx0=(Vel_fct(ii))*(rrX(ii,:))
          Vx=Vx0
          xx=xx*0
          xx(1)=-R
          drX=(sqrt(sum(xx**2)))*(0.99)

          Xstore=Xstore*0
          Vstore=Vstore*0
          Fx_Stored=Fx_Stored*0
          ct=0
          do while ((abs(drX/R))<1.0)
            ct=ct+1
            xx=xx+(Vx*dt)
            drX=(sqrt(sum(xx**2)))

            do jj=1,3
              if (xx(jj)==0) then
                dir(jj)=0
              else
                dir(jj)=-(xx(jj)/(abs(xx(jj))))
            endif
          enddo
          Fx=(abs(F_VDW_m*((xx/R)**3)))*dir
          do jj=1,3
            Vx(jj)=Vx(jj)+(Fx(jj)*dt/m_m)
          enddo
          write(ii0,*) xx(:),Vx(:)

          do jj=1,3
            Xstore(ct,jj)=xx(jj)
            Vstore(ct,jj)=Vx(jj)
            Fx_Stored(ct,jj)=Fx(jj)
          enddo
          if (ct>(dx0*10)) then
            drX=10*R
            print *,'PROBLEM!!!',ct,dx0
          endif

          enddo

          Tct(ii)=ct
          Total_CT0=Total_CT0+ct
          Total_CT=Total_CT+ct
          Vel_travel=Vel_travel*0
          do jj=1,3
            fooreal1=0
            do kk=1,ct
              fooreal1=fooreal1+(Vstore(kk,jj)/ct)
            enddo
            Vel_travel(jj)=fooreal1
          enddo
          do jj=1,3
            Vel_travelDat(ii)=Vel_travelDat(ii)+(Vel_travel(jj)**2)
          enddo
          Vel_travelDat(ii)=sqrt(Vel_travelDat(ii))


          VelF=0
          do jj=1,3
            VelF=VelF+(Vstore(ct,jj)**2)
          enddo
          VelF=sqrt(VelF)
          VelFdat(ii)=VelF

          Vrat=Vstore(ct,:)/VelF
          Xrat=xx/R

          VXdot=0
          do jj=1,3
            VXdot=VXdot+(Xrat(jj)*Vrat(jj))
          enddo
          VXrat(ii)=VXdot

c ------- Calculate the force, to numerically determine the pressure
          F_dat(ii)=((2*m_m*VXdot)*VelF/(ct*dt))-F_VDW_m

        enddo
        close(ii0)
        Total_Ct_Fct(ii0)=Total_Ct0
        Total_Ct0=0
      enddo
      
c --- Calculate the average and RMS position and velocity of the molecules

      Avg6=Avg6*0.
      RMS3=RMS3*0.
      StDev6=StDev6*0.

      do jj=1,ctsplit

        if (jj<10) then
          write(filenameSV,'("SaveXV3_00",I1,".txt")')jj
        elseif (jj<100) then
          write(filenameSV,'("SaveXV3_0",I2,".txt")')jj
        elseif (jj<1000) then
          write(filenameSV,'("SaveXV3_",I3,".txt")')jj
        endif
        open(unit=jj,file=filenameSV)

        do ii=1,(Total_Ct_Fct(jj))
          read(jj,*) Read6(:)
          Avg6=Avg6+Read6
          RMS3=RMS3+(Read6(4:6)**2.)
        enddo
        close(jj)
      enddo
      Avg6=Avg6/Total_CT
      RMS3=sqrt(RMS3/Total_CT)

c --- Calculate the standard deviation position and velocity of the molecules

      do jj=1,ctsplit

        if (jj<10) then
          write(filenameSV,'("SaveXV3_00",I1,".txt")')jj
        elseif (jj<100) then
          write(filenameSV,'("SaveXV3_0",I2,".txt")')jj
        elseif (jj<1000) then
          write(filenameSV,'("SaveXV3_",I3,".txt")')jj
        endif
        open(unit=jj,file=filenameSV)

        do ii=1,(Total_Ct_Fct(jj))
          read(jj,*) Read6(:)
          StDev6=StDev6+((Read6-Avg6)**2)
        enddo
        close(jj)
      enddo
      StDev6=StDev6/Total_CT





      
c --- Output results of this specific trial

      P_kinetic=0
      U_KE=0
      S_entropy=0
      do ii=1,Np
            P_kinetic=P_kinetic+(F_dat(ii))
            U_KE=U_KE+(VelFdat(ii)**2)
            S_entropy=S_entropy+(Vel_travelDat(ii))
      enddo

      P_kinetic=((P_kinetic/Np)*(Av/AreaS))*(R/Rb)
      P_IG=Av*Kb*T/V

      U_KE=((U_KE/Np))*(0.5*Av*m_m)
      U_PE=-dP_VDW*V
      U=U_KE+U_PE
      S_entropy=((3*log(S_entropy/Np))+(log(V-(b*MM))))*Av*Kb

      outputdat(1)=U
      outputdat(2)=S_entropy
      outputdat(3)=P_kinetic
      outputdat(4)=P_PR
      outputdat(5:10)=Avg6
      outputdat(11:16)=StDev6
      outputdat(17:19)=RMS3
      outputdat(20)=Total_CT
      outputdat(21)=V_rms_calc
      outputdat(22)=V_avg_calc

      end subroutine

c ==========================================================


c ==========================================================

c --- Subroutine to calculate intermolecular attractive force component
c --- Equation developed to match Peng-Robinson equation of state

      subroutine Get_dP_VDW(Vel,Vr,dP_VDW)

      real, intent(in) :: Vr
      double precision, intent(in) :: Vel
      double precision, intent(out) :: dP_VDW
      double precision pi,Kb,Av,m_m,T_eff,Tr,Pc
      double precision MM,Tc,Vc,V,R,Coeff,Rb,bp,aa
      integer ii

      pi=3.1415926535897932384626
      Kb=1.38064852e-23 
      Av=6.02214086e23

c     Argon
      MM=.0399  ! Molar Mass (kg/mole)
      Pc=4.863e6  ! Critical Pressure (Pa)
      Tc=150.687  ! Critical Temperature (K)
      Vc=1./535    ! Critical Volume (m^3/kg)
c      ecc=0       ! Eccentricity factor

      V=Vr*Vc*MM

      m_m=MM/Av
      Rg=Av*Kb/MM

      bp=(2.**(1./3))-1.
      aa=(1/(9*bp))*(Rg**2)*(Tc**2.5)/Pc

      T_eff=(Vel**2)*m_m/(3*Kb)
      Tr=T_eff/Tc


      if (Tr<1.) then
        Coeff=(2.3246+(-0.8441/(sqrt(Vr)))+(-0.8670))*Tr
      else
        Coeff=2.3246+(-0.8441/(sqrt(Vr)))+(-0.8670*sqrt(Tr))
      endif
      if (Coeff>1) then
        Coeff=0
      endif

      dP_VDW=(aa/(sqrt(T_eff)))/((V/MM)**2)
      dP_VDW=dP_VDW*Coeff

      
      end subroutine

c ==========================================================

c ==========================================================


c --- Subroutine to distribute velocities of molecules
c --- Ensures proper ratio of average and RMS velocity

      subroutine make_rand_fct(NN,randdat)

      integer, intent(in) :: NN
      double precision, intent(out) :: randdat(NN)

      double precision , allocatable :: NormFct(:),Xfct(:)
      double precision , allocatable :: NormFct0(:)
      double precision ctX,x,MinX,stdev0,dx,foo
      integer ii

c      ALLOCATE(randdat(NN))
      ALLOCATE(Xfct(NN))
      ALLOCATE(NormFct0(NN))
      ALLOCATE(NormFct(NN))

      MinX=0.200
      stdev0=0.71

      dx=((1.-MinX)*2.)/(NN-1)

      ctX=0.0
      do ii=1,NN
        x=MinX+((ii-1)*dx)
        Xfct(ii)=x
        NormFct(ii)=(exp(-0.5*(((x-1)/stdev0)**2)))
        ctX=ctX+(1./(exp(-0.5*(((x-1)/stdev0)**2))))
      enddo

      randdat(1)=MinX
      do ii=2,NN
        foo=(((1.-MinX)*2.)*(1./NormFct(ii))/ctX)
        randdat(ii)=randdat(ii-1)+foo
      enddo



      end subroutine

c ==========================================================

c ==========================================================

\end{verbatim}

\end{document}